\documentclass[
prd,
superscriptaddress,
tightenlines,
showpacs,
showkeys,
nofootinbib,
aps,
amsfonts,
amssymb,twocolumn
]{revtex4}

\usepackage[usenames,dvipsnames]{xcolor}
\usepackage{amsmath,mathrsfs}
\usepackage{amsmath}
\usepackage{amsfonts}
\usepackage{amssymb} 
\usepackage{graphicx}
\usepackage{epic}
\usepackage{eepic}
\usepackage{epsfig}
\usepackage{latexsym}
\usepackage{float}
\usepackage{shorthand}
\usepackage{multirow,array}
\usepackage[caption=false]{subfig}

\usepackage[
    colorlinks,
    linkcolor=blue,           
    citecolor=red,            
    filecolor=magenta,        
    urlcolor=magenta,          
    hyperfootnotes]{hyperref}

\renewcommand{\lq}{\ell_q}
\newcommand{\ubr}[1]{\raisebox{1.5ex}{\hspace{#1ex}$\frown$\relax}}
\newcommand{\lbr}[1]{\raisebox{-1.5ex}{\hspace{#1ex}$\smile$\relax}}

\begin{document}

\title{{\hfill{\rm\small HRI-RECAPP-2015-003, MITP/15-015}\\~\\}Single Productions of Colored Particles at the LHC: An Example with Scalar Leptoquarks}

\author{Tanumoy Mandal}
\email{tanumoymandal@hri.res.in}
\affiliation{Regional Centre for Accelerator-based Particle Physics, Harish-Chandra Research Institute, Chhatnag Road, Jhusi, Allahabad - 211019, India}
\author{Subhadip Mitra}
\email{subhadipmitra@gmail.com}
\affiliation{Department of Physics, IIT Kanpur, Kanpur 208016, India}
\author{Satyajit Seth}
\email{sseth@uni-mainz.de}
\affiliation{PRISMA Cluster of Excellence, Institut f\"{u}r Physik,
Johannes Gutenberg-Universit\"{a}t Mainz, 
D\,-\,55099 Mainz, Germany}


\begin{abstract}
Current LHC searches for new colored particles generally focus on their pair production channels and assume any single production to be negligible. We argue that such an assumption may be unnecessary in some cases. Inclusion of model dependent single productions in pair production searches (or vice versa) can give us new information about model parameters or better exclusion limits. Considering the example of the recent CMS search for first generation scalar leptoquarks in the pair production channel, we illustrate how single productions can be systematically included in the signal estimations and demonstrate how it can affect the mass exclusion limits and give new bounds on leptoquark-lepton-quark couplings. We also estimate the effect of the pair production in the more recent CMS search for scalar leptoquarks in single production channels.
\end{abstract}



\maketitle 

\begin{figure*}[!t]
        \centering
		\subfloat[]{
                \includegraphics[width=0.18\linewidth]{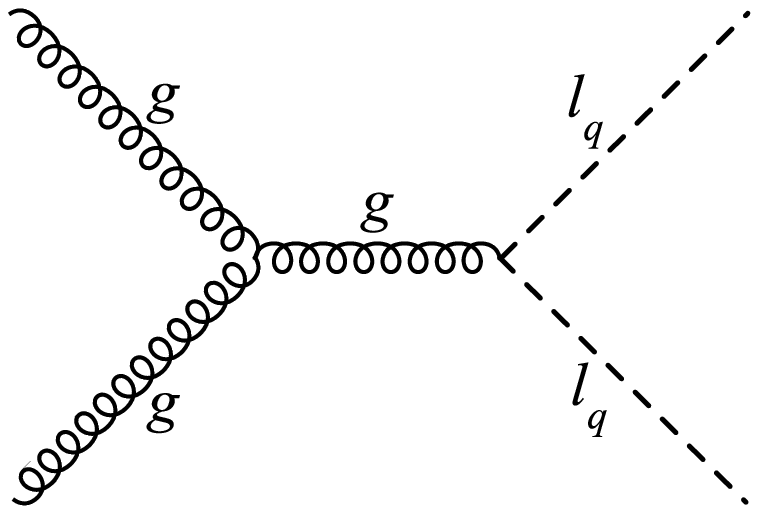}
                \label{fig:gg2SS1}}
		\hfill{}
        \subfloat[]{
                \includegraphics[width=0.18\linewidth]{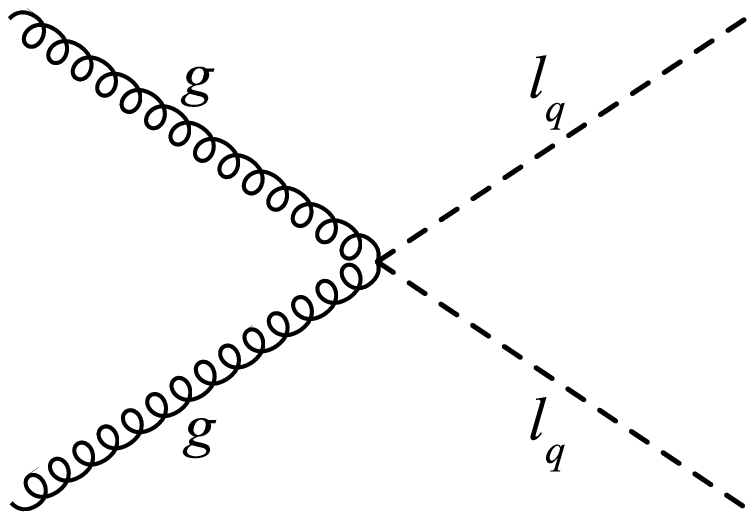}
                \label{fig:gg2SS3}
                }
		\hfill{}
        \subfloat[]{
                \includegraphics[width=0.18\linewidth]{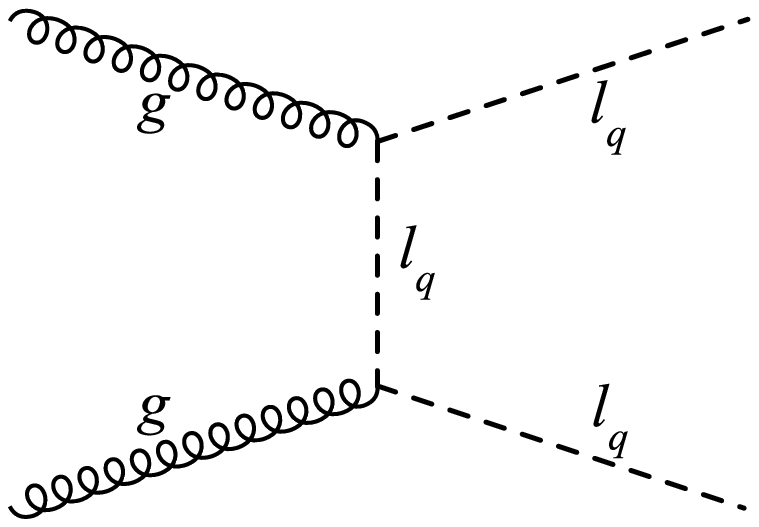}
                \label{fig:gg2SS2}
                }
		\hfill{}
		\subfloat[]{
                \includegraphics[width=0.18\linewidth]{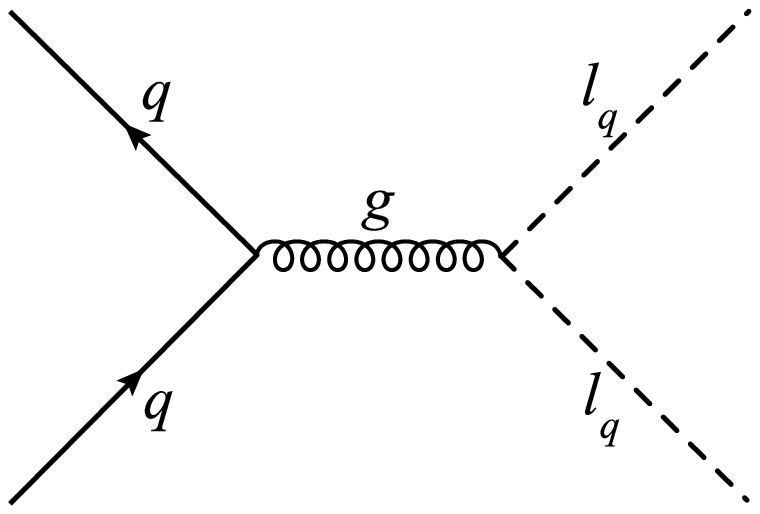}
                \label{fig:qq2SS1}
        			}
		\hfill{}
		\subfloat[]{
                \includegraphics[width=0.18\linewidth]{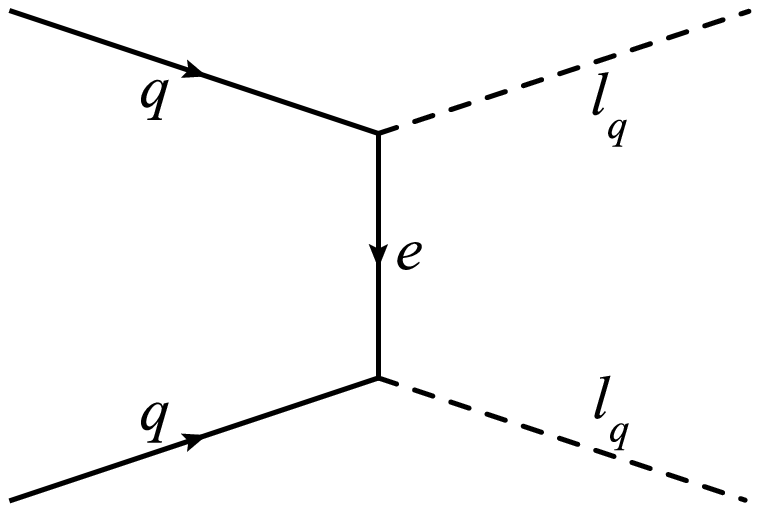}
                \label{fig:uu2SS2}
        			}\\
		\subfloat[]{
                \includegraphics[width=0.18\linewidth]{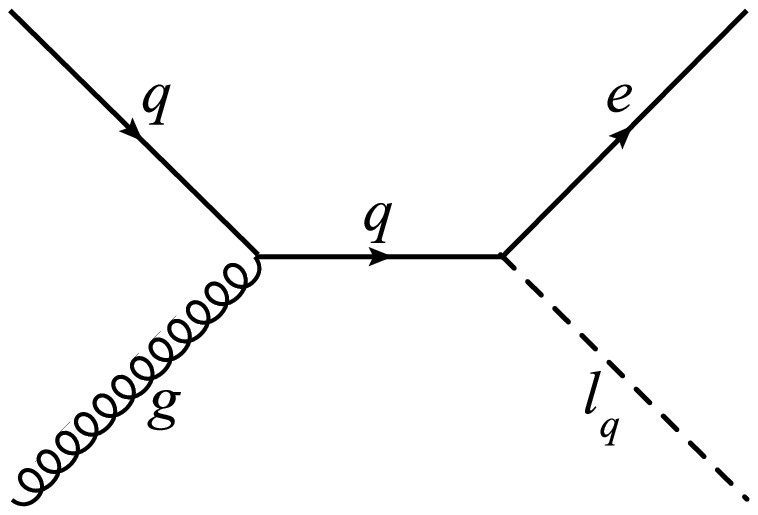}
                \label{fig:ug2Se1}}
		\hfill{}
        \subfloat[]{
                \includegraphics[width=0.18\linewidth]{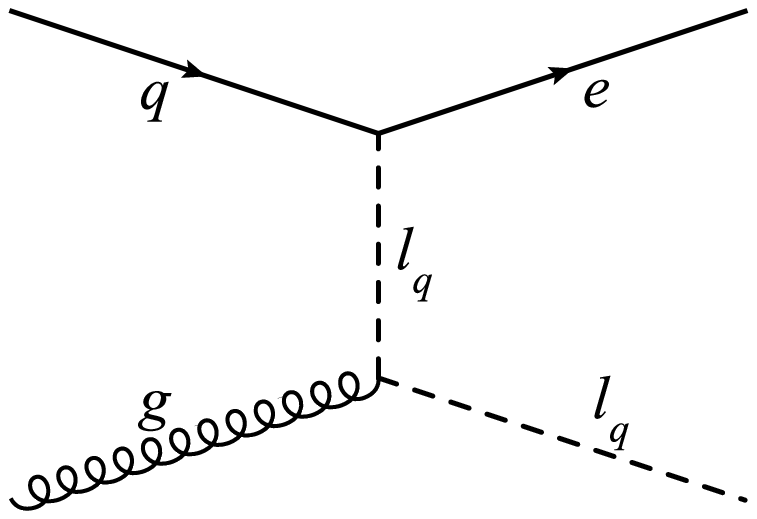}
                \label{fig:ug2Se2}
                }
		\hfill{}
        \subfloat[]{
                \includegraphics[width=0.18\linewidth]{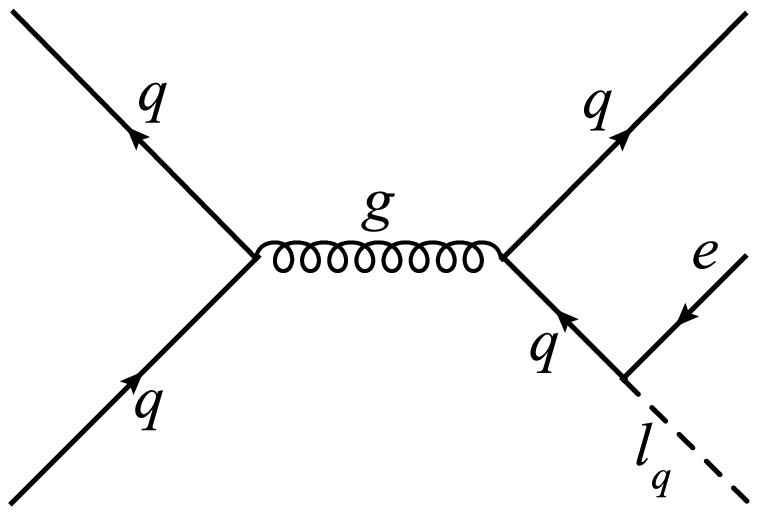}
                \label{fig:qq2Sej1}
                }
		\hfill{}
		\subfloat[]{
                \includegraphics[width=0.18\linewidth]{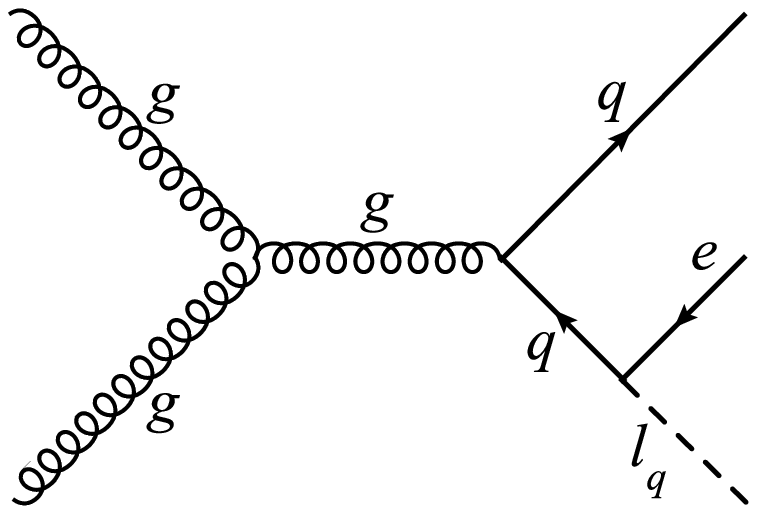}
                \label{fig:gg2Sej1}
        			}
		\hfill{}
		\subfloat[]{
                \includegraphics[width=0.18\linewidth]{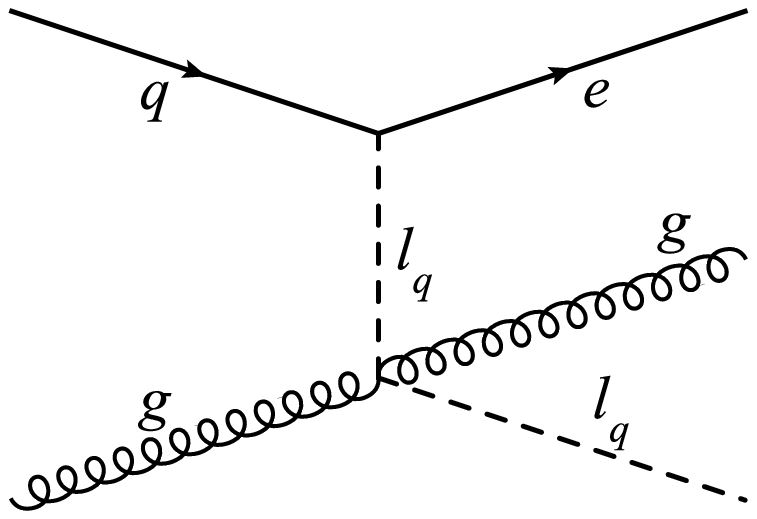}
                \label{fig:ug2Sej1}
        			}
        \caption{Representative parton level Feynman diagrams for pair [(a)-(e)] and  single productions [(f)-(j)] of LQs at the LHC. }\label{fig:lqFDs}
\end{figure*}

\section{Introduction}\label{sec:intro}

Various beyond the Standard Model (BSM) theories predict a host of new heavy particles at the TeV range. The Large Hadron Collider (LHC) at CERN is searching for many of these particles. In a hadron collider like the LHC, a heavy colored particle could be produced via different mechanisms. Among them, usually the strong interaction mediated pair production dominates. Other direct production processes like single productions generally depend on some model specific couplings  ($\lm_{\rm x}$). 
Hence, many present direct searches for new colored particles at the LHC (and related phenomenological studies) focus on their pair productions and ignore single productions assuming the couplings controlling them to be smaller than the strong coupling. With this assumption the results (or predictions) seemingly become model independent. But this, however, raises a question -- is it justified to ignore single production completely? Obviously, the question is relevant if any such ignored $\lm_{\rm x}$ is indeed not very small in the nature. But, even if $\lm_{\rm x}$ is small, cross-sections of some single productions could still be comparable to the pair production cross-section in some region of parameter-space. This is possible when the particle being produced is quite heavy so that the extra phase-space suppression received by the  pair production is significant. Hence, at present when no signatures of these new particles have been found and, as a result, the direct search experiments are regularly pushing the lower limit of the allowed masses of these new particles upwards, the question gets increasingly important.

In this paper, we want to illustrate that assuming single productions to be negligible compared to pair production may be unnecessary (or even improper) in some cases. In fact, including single productions in the theory estimation might give us extra clues about the BSM models being probed. We shall use the example of the recent leptoquark (LQ) search by CMS \cite{CMS:2014qpa} to indicate how single productions can be systematically included in this search for pair production of scalar LQs and demonstrate how the extracted exclusion limits (ELs) on parameters (like masses, branching fractions (BFs) etc.) change in presence of model dependent production processes with similar final states as the pair production which could potentially `contaminate' the signal simulations.\\

LQs are hypothetical color-triplet bosons (scalars or vectors) that also carry lepton quantum numbers. Hence, a heavy LQ can decay to a lepton and a quark. They appear in different BSM theories like the Pati-Salam models \cite{Pati:1974yy}, SU(5) grand unified theories \cite{Georgi:1974sy},   models with quark lepton compositeness (see e.g. \cite{Schrempp:1984nj}), colored Zee-Babu model \cite{Kohda:2012sr} etc. In some supersymmetric models squarks can also couple to a quark and a lepton via some R-parity violating couplings \cite{Barbier:2004ez} and thus have similar phenomenology to the LQs. 
The leptons in their decay states make their collider signature relatively clean compared to that of other colored particles that dominantly decay hadronically. At the LHC, searches for LQs have been going on for quite some time. Both ATLAS and CMS have put limits on LQ masses \cite{CMS:2014qpa,Aad:2011ch,ATLAS:2012aq,ATLAS:2013oea,Chatrchyan:2012vza,CMS:zva,Khachatryan:2014ura,CMS:2015kda}.

Recently, CMS collaboration has reported the results of their searches for the first generation scalar LQs at the 8 TeV LHC with 19.6 fb$^{-1}$ of integrated luminosity \cite{CMS:2014qpa}. They have searched for pair production of LQs in two different channels -- i) the $eejj$ channel where both LQs decay to  electrons and jets and ii) the $e\nu jj$ channel where only one LQ decays to an electron and a jet while the other one decays to a neutrino and a jet. They have found a mild $2.4\sg$ ($2.6\sg$) excess of events compared to the Standard Model (SM) background in the $eejj$ ($e\nu jj$) channel for LQs with mass ($M_{\lq}$) around 650 GeV. The excesses observed certainly make the LQ search more interesting to be investigated further \cite{Queiroz:2014pra,Allanach:2015ria,Allanach:2014nna,Chun:2014jha,Bai:2014xba,Varzielas:2015iva}.
The analysis also puts 95\% CL ELs on first generation scalar LQs for $M_{\lq} < 1005\ (845)$ GeV for $\bt = 1\ (0.5)$, where $\bt$ is the BF for a LQ to decay to an electron-quark pair.

Here, we take a closer look at the CMS analysis and investigate LQ single production processes that could give potentially significant contribution  to the signal \cite{Belyaev:2005ew} but were ignored in the analysis. In the next section, we specify the single production processes under consideration, in Sec. \ref{sec:sig_sim} we present the phenomenological LQ models and discuss the method that we use to compute single productions, in Sec. \ref{sec:results} we present our results and finally, in the light of the new results we further clarify our argument and conclude in Sec. \ref{sec:conclusion}.

\section{LQ single productions at the LHC}\label{sec:sing_prod}

The limits from HERA data \cite{Collaboration:2011qaa,Abramowicz:2012tg} roughly indicate that the generic (scalar) LQ-lepton-quark coupling, $\lm$ could even be larger than 0.5 if $M_{\lq}\gtrsim 600$ GeV.  However, the CMS analysis assumes that $\lm$ is small enough to ignore all LQ single productions and sets $\lm=\lm_{\rm QED}=0.3$ to compute the total decay width of a scalar LQ for the signal simulations. In the leading order (LO) of couplings, the pair production of LQs is almost independent of $\lm$ (see Figs. \ref{fig:gg2SS1} - \ref{fig:qq2SS1}) except for the $t$-channel lepton exchange contribution whose amplitude is proportional to $\lm^2$ (Fig. \ref{fig:uu2SS2}). For small $\lm$, this contribution is actually quite small, e.g. with $\lm=0.3$ it amounts to only about 5 percent of the $\lm$ independent contribution to the LO pair production cross-section for $M_{\lq}=650$ GeV (see Table \ref{tab:sigma}). This is why it is justified when the $\lm^2$ contribution is ignored in Refs. \cite{Kramer:2004df,Mandal:2015lca} to compute the next-to-leading order (NLO) cross-sections for the pair production (used in the CMS analysis). However, while it may be safe to ignore the $t$-channel lepton exchange contribution to the pair production for $\lm=0.3$, we shall see that even with this value of $\lm$, single production of LQs could actually contribute quite significantly to the analysis.

The important point here is that a fraction of the single production events would  pass the signal selection cuts used in the CMS analysis and contribute to their estimations of pair production events. For example, in the $eejj$ channel, the analysis considers $pp\to (\lq\,\lq)\to e j \ubr{-2.5}\, ej\lbr{-2.}$ (the curved connections above or below mark a pair coming from the decay of  a leptoquark, $\lq$) and hence demands the presence of exactly two electrons and at least two jets in the selection. However, for non-zero $\lm$, any $pp\to (\lq ej)\to ej\ubr{-2.5}\,ej$ process could also contribute to the signal. Hence, \emph{a priori}, these should also be included in the signal simulations. 
Since the CMS analysis requires at least two jets for event selections, the following two types of hard processes with 3-body final state can contribute:
\begin{enumerate}
\item  Born single production with radiation (BR1):   The Born diagrams  for $pp\to \lq \ell$ (as shown in Figs. \ref{fig:ug2Se1} \& \ref{fig:ug2Se2}) can not directly contribute to the CMS search but with the emission of a QCD radiation (we refer to both parton splitting and radiation simply as `radiation') they contribute at $\mc O\lt(g_s^2\lm\rt)$.

\item New subprocess of three-body single production (NS3): It originates at {${\mathcal{O}}(g_s^2\lm)$} and includes diagrams for $pp \to \lq\ell j$ that do not count as BR1 single production (see Figs. \ref{fig:qq2Sej1}-\ref{fig:ug2Sej1} for example).
\end{enumerate}

\section{LQ model and signal simulations}\label{sec:sig_sim}

As already stated, CMS assumes that a generic first generation scalar LQ decays to an electron and a quark with a BF $\bt$ and a neutrino and a quark rest of the time. Hence, we also adopt a parametrization in which a LQ decays to a $eq$ pair via a coupling $\lm_{e}$ with BF$_{\lq\to eq}=\bt_e=\bt$ and to a $\n q$ pair via $\lm_{\n }$ with BF$_{\lq\to \n q}=\bt_\n=(1-\bt)$, and take
\ba
\lm_{e}^2 = \bt_e \lm^2=\bt\lm^2,\quad  \lm_{\n }^2 = \bt_\n \lm^2=\lt(1-\bt\rt)\lm^2\;, 
\ea
so that for any $\bt$, the total decay width remains fixed at, 
\ba 
\Gamma_{\lq} = \frac{\lm_e^2}{16\pi}\,M_{\lq} + \frac{\lm_\n^2}{16\pi}\,M_{\lq} = \frac{\lm^2}{16\pi}\,M_{\lq},
\ea
i.e., the expression used in the CMS analysis. Motivated by the experiment, we consider scalar LQs that couple with only first generation quarks and take two simple Lagrangians for scalar LQs with electromagnetic charge, $Q_{\rm EM}=-1/3$ (Model A) and $2/3$ (Model B) respectively,
\begin{eqnarray}
\mathscr L_{\rm int}^{\rm A} &=& \lm_{e} \left(\sqrt{\eta_{\rm L}}\, \bar u_{\rm R} e^+_{\rm L} + \sqrt{\eta_{\rm R}}\, \bar u_{\rm L} e^+_{\rm R}\right)\lq+ \lm_{\nu } \bar d_{\rm R} \tilde \nu_{\rm L}^e \lq\nn\\ 
&& +\ {\rm H.c.},\label{eq:modelA}\\
\mathscr L_{\rm int}^{\rm B} &=& \lm_{e} \left(\sqrt{\eta_{\rm L}}\, \bar d_{\rm R} e^-_{\rm L} + \sqrt{\eta_{\rm R}}\, \bar d_{\rm L} e^-_{\rm R}\right)\lq + \lm_{\nu } \bar u_{\rm R} \nu_{\rm L}^e \lq\nn\\
&& +\ {\rm H.c.},\label{eq:modelB}
\end{eqnarray}
where $\eta_{\rm L}$ and $\eta_{\rm R}=(1-\eta_{\rm L})$ are the electron chirality fractions, i.e., $\eta_{\rm L} \left(\eta_{\rm R}\right)$ gives the fraction of electrons coming from a LQ decay that are left-handed (right-handed). In other words, $\eta_{\rm L}\bt_e \left(\eta_{\rm R}\bt_e\right)$ is the BF for a LQ to decay into a left-handed (right-handed) electron and a quark. As the experiment is insensitive to the polarization of electrons, for simplicity and without loss of any generality, we have set $\eta_L$ = 1 for our computation.

The above models are completely generic and as long as no distinction is made between $e^+$ and $e^-$, they can act as templates for a wider variety of LQs that can decay to SM quarks and leptons. For example, an analysis done in Model A for the $eejj$ channel separately will also be  applicable for a LQ with charge $5/3$ that couples to a $u$-type quark and an electron (but not to a neutrino). Similarly, the LQ from Model B can represent a charge $-4/3$ particle also. Moreover, these models could be connected to the ones generally found in literature (see e.g., \cite{Blumlein:1992ej,Hewett:1997ce}) via parameter rescaling. 

With these models, we have performed Monte Carlo (MC) simulations for the LQ single productions in both $eejj$ and $e\n jj$ channels (as long as the LQ width remains small, we ignore the interference between single and pair production). For our simulations we have employed \textsc{MadGraph5} \cite{Alwall:2014hca} (and \textsc{Pythia}6 \cite{Sjostrand:2006za} within it) to generate and shower events and \textsc{Delphes} 3 \cite{deFavereau:2013fsa} to simulate the CMS detector environment and implement the selection cuts. We have used CTEQ6L1 Parton Distribution Functions (PDFs) \cite{Pumplin:2002vw} for all our numerical computations.

\begin{table*}[!t]\caption{Cross-sections (in fb) for LQ productions at the 8 TeV LHC. The cross-sections are obtained with $\m_{\rm R}=\m_{\rm F}=M_{\lq}$. 
The inclusive single production cross-sections, $\sg_{\rm s}^{ee}(\bt)=\bt_e\times\sg(pp\to\lq e + n j)\approx\lm^2\bt^2\bar\sg_{\rm s}^{ee}$ (see Eq.~\ref{eq:sg_ee_bt_dep}) and $\sg_{\rm s}^{e\nu}(\bt) = \bt_\n\times\sg(pp\to\lq e+ n j)+\bt_e\times\sg(pp\to \lq \nu +n j)\approx 2\lm^2\bt(1-\bt)\bar\sg_{\rm s}^{e\nu}$ (see Eq.~\ref{eq:sg_ev_bt_dep}) with $n\geq0$ are obtained by ME$\oplus$PS technique (as described in Sec~\ref{sec:eejj}) with Model A (Eq.~\ref{eq:modelA}).}\label{tab:sigma}

\begin{tabular}{|c|ccc|cc|cc|}
\hline 
LQ&\multicolumn{3}{c|}{Pair production}&\multicolumn{4}{c|}{Inclusive single production}\\ \cline{5-8}
Mass&\multicolumn{3}{c|}{$\sg^{\rm LO}_{\rm p}(pp\to \lq\lq)$ }&\multicolumn{2}{c|}{$\sg_{\rm s}^{ee}(\bt=1.0)$ }&\multicolumn{2}{c|}{$\sg_{\rm s}^{e\nu}(\bt=0.5)$}
\\
 (GeV)&$\lm\to0$ &$\lm=0.3$&$\lm=0.5$&$\lm=0.3$&$\lm=0.5$&$\lm=0.3$&$\lm=0.5$\\
 \hline\hline
550	& 24.4		&22.4	&21.8	&41.2	&118.7	&17.0	&48.2	\\
650	& ~7.2	    &~6.6	&~6.4 	&17.7	&~51.3 	&~7.3	&20.5	\\
750	& ~2.4		&~2.1	&~2.1	&~8.3	&~24.5 	&~3.5	&~9.7	\\\hline

\end{tabular}
\end{table*}
\begin{table*}[!t]\caption{MC events for LQ signals obtained for the 8 TeV LHC and 19.6 fb$^{-1}$ of integrated luminosity. 
For the pair production, we quote the number of MC signal events obtained by CMS, $\mc N_{\rm p}$ \cite{CMS:2014qpa} (with $\sg^{\rm NLO}_{\rm p}$\cite{Kramer:2004df}). For single productions, we show the number of events obtained with Model A (Eq.~\ref{eq:modelA}) that survive the same final $eejj$ or $e\nu jj$ selection criteria as the pair production for each $M_{\lq}$ (described in the CMS report), $\mc N_{\rm s}$. We also show the selection-cut efficiency, $\ep^{ee/e\n}_{\rm s}(M_{\lq})$ -- the fraction of single production events that survives the selection criteria for a particular $M_{\lq}$ (see Eq. \ref{eq:efficiency_single}).}\label{tab:evntscms}

\begin{tabular}{|c|c|cc|c|cc|}
\hline 
LQ&\multicolumn{3}{c|}{$eejj$ channel $(\lm=0.3,\bt=1.0)$}&\multicolumn{3}{c|}{$e\nu jj$ channel $(\lm=0.3,\bt=0.5)$}\\ \cline{2-7}
Mass&Pair prod.  &\multicolumn{2}{c|}{Single prod.}&Pair prod. &\multicolumn{2}{c|}{Single prod.}\\ 
(GeV)&Events ($\mc N_{\rm p}$) & Events ($\mc N_{\rm s}^{ee}$) & $\ep_{\rm s}^{ee}$ (\%) &Events ($\mc N_{\rm p}$) & Events ($\mc N_{\rm s}^{e\n}$)& $\ep_{\rm s}^{e\n}$ (\%)\\
 \hline\hline
550	& 410.5$\pm$1.9	   &83.0		&10.3		& 121.4$\pm$1.2	     &18.1	&~5.4	\\
650	& 125.9$\pm$0.6	   &30.8		&~8.9		& ~37.2$\pm$0.4		 &~6.2	&~4.3	\\
750	& ~43.1$\pm$0.2    &12.4		&~7.6		& ~12.9$\pm$0.1   	 &~2.6	&~3.8	\\\hline

\end{tabular}
\end{table*}

\subsection{The $eejj$ channel}\label{sec:eejj}

Following the CMS analysis, we set $\bt=1$ to generate the signal events from the single productions in the $eejj$ channel. 
Unlike the pair production, events from the lowest order single production process (strictly Born diagrams) will not pass the final event selections as they will have only one jet in the final state. At the parton level, one could roughly estimate the (potential) contribution to the $eejj$ channel coming from the inclusive single production ($\sg^{ee}_{\rm s}$) by computing  the cross-section for
\ba
pp\to\ (\lq e j) \ \to ej\ubr{-2.5}\, e j 
\ea
without including the pair production diagrams. This will include contributions from both BR1 and NS3 single production processes. But the BR1 process contains divergent pieces that get cancelled when loop diagrams (virtual correction to Born diagrams) are properly included. The picture becomes clear if we express the cross-section for the inclusive single production  with two electrons in the final state as,
\begin{widetext}
 \ba 
 \sg^{ee}_{\rm s}
= \underbrace{\big(\sg^{\rm  LO}_{(\lq e)} + \overbrace{ \sg^{\rm  virtual}_{(\lq e)} + \sg_{{\rm BR1}(\lq ej)}^{\rm  soft+collinear}}^{\rm Divergent~terms} + \sg_{{\rm NS3}(\lq ej)}^{\rm  soft}\big)}_{\rm Small~contribution}
\ +\ 
\underbrace{\big(\sg_{{\rm BR1}(\lq ej)}^{\rm  hard} + \sg_{{\rm NS3}(\lq ej)}^{\rm hard}\big)}_{\rm Main~contribution}\ +\ \cdots \;,\label{eq:softhard}
\ea
\end{widetext} 
where the neglected terms are of higher order than $\mc O\lt(\al_s^2\lm^2\rt)$ and the superscripts `soft + collinear', `soft' or `hard' refer to the nature of the extra jet.
Notice that, just like the `pure' Born process, the pieces with divergence ultimately contribute very little to the experiment (if only parton showers are included)  because it demands two separated `hard' jets (see the selection criteria in \cite{CMS:2014qpa}). Hence, in a tree level calculation, one could put some minimum $p_T$-cut on the second jet (the jet coming from the LQ is generally the hardest one) to avoid the small-contributing part. This way one could compute the cross-section, but it would then become a function of the $p_T$-cut and hence ambiguous (since \emph{a priori} there is no absolute way to choose the `ideal' $p_T$-cut) and unsuitable for our purpose.

Clearly, a better approach would be to consider loop diagrams consistently to compute the $pp\to \lq e$ processes upto NLO QCD level (keeping the hard radiation) including the NS3 contribution to compute the inclusive single production cross-section ($\sg^{ee}_{\rm s}$) correctly upto $\mc O\left(\al_s^2\lm^2\right)$ without any ambiguity.  We are at present working in this direction \cite{MMSLQSNLO}\footnote{Although not directly useful for this paper, we note that some estimates of the NLO LQ single productions at the 14 TeV LHC in different models already exist in the literature (see e.g. \cite{Alves:2002tj,Hammett:2015sea}).} but for this paper we employ matrix element-parton shower matching (ME$\oplus$PS) technique to obtain a theoretical estimate of the inclusive single production cross-section without any arbitrary $p_T$ cut on the second jet. The main difference between a proper NLO QCD computation and our ME$\oplus$PS computation will come from the part that contributes very little (from the second and third terms of Eq. \ref{eq:softhard}). Hence, for the same luminosity, the number of events passing the selection criteria (require atleast two jets with $p_T\geq45$ GeV) \cite{CMS:2014qpa} should not vary too much between a proper loop included NLO level computation and the one with ME$\oplus$PS. The estimates for the inclusive cross-section will be different but this difference should not matter if the efficiencies of the selection cuts (will be defined in Sec. \ref{sec:results}) are computed consistently.

For the ME$\oplus$PS computation we have used the shower-$k_T$ scheme \cite{Alwall:2008qv}. We generate events for the inclusive single  $\lq$ production signal as the combination of the following processes,
\be
\left. \begin{array}{lclcl}
pp &\to &(\lq\ e) 		&\to & e j\ubr{-2.5}\ e\,,\\
pp &\to &(\lq\ ej)		&\to & e j\ubr{-2.5}\ ej\,,\label{eq:matching_ee}\\
pp &\to &(\lq\ ejj)  	&\to & e j\ubr{-2.5}\ ejj\,.
\end{array}\right\}
\ee The $\lm$ and $\bt$ dependence of inclusive single production cross-section in the $eejj$ channel can be made explicit as:
\ba
\sg^{ee}_{\rm s}\lt(\bt,\lm,M_{\lq}\rt) &=&  \lm_e^2\ \bt_e\ \bar\sg^{ee}_{\rm s}\left(M_{\lq}\right) + \cdots\nn\\
&\overset{\rm def.}{=}& \lm^2\ \bt^2\ \bar\sg^{ee}_{\rm s}\left(M_{\lq}\right)+ \cdots\,,\label{eq:sg_ee_bt_dep}
\ea
where the neglected terms are at least $\mc O(\lm^4)$. Notice that with the chosen parametrization, the $\bt$ dependence of the single production cross-section becomes the same as the pair production.

\begin{figure*}[!t]
\begin{tabular}{rm{0.4\linewidth}m{0.08\linewidth}m{0.4\linewidth}l}
&\subfloat[ $eejj$ channel (taken from \cite{CMS:2014qpa})]{\includegraphics[width=\linewidth]{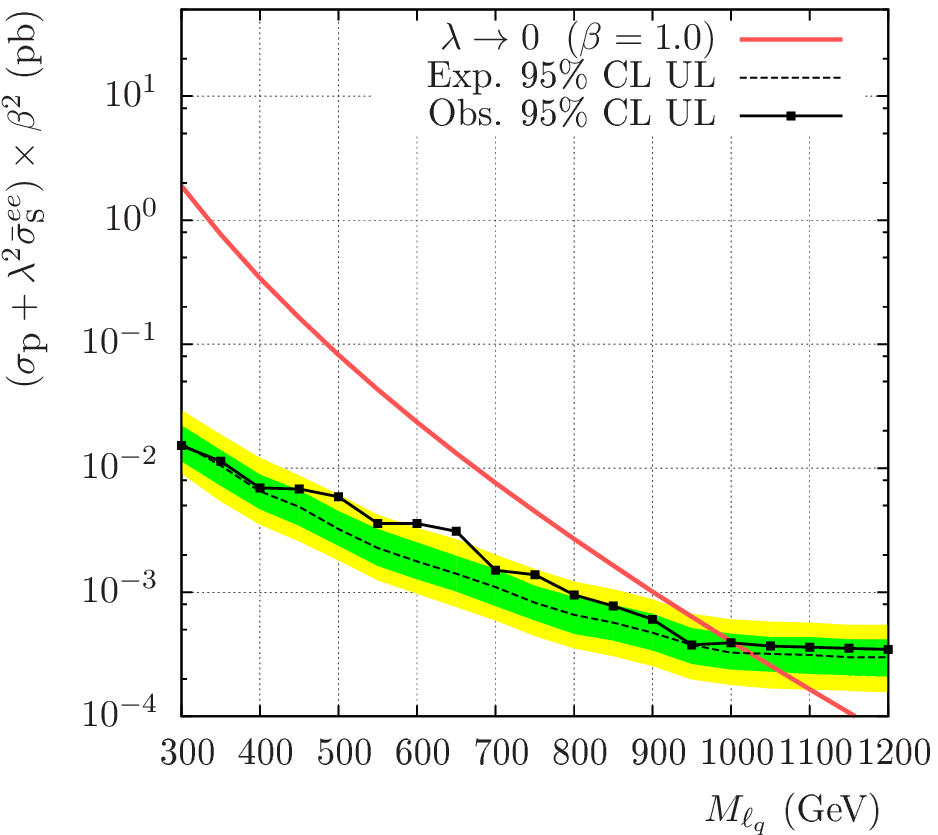}\label{fig:sig_mass_eejj_CMS}} 
&&
\subfloat[$e\n jj$ channel (taken from \cite{CMS:2014qpa})]{\includegraphics[width=\linewidth]{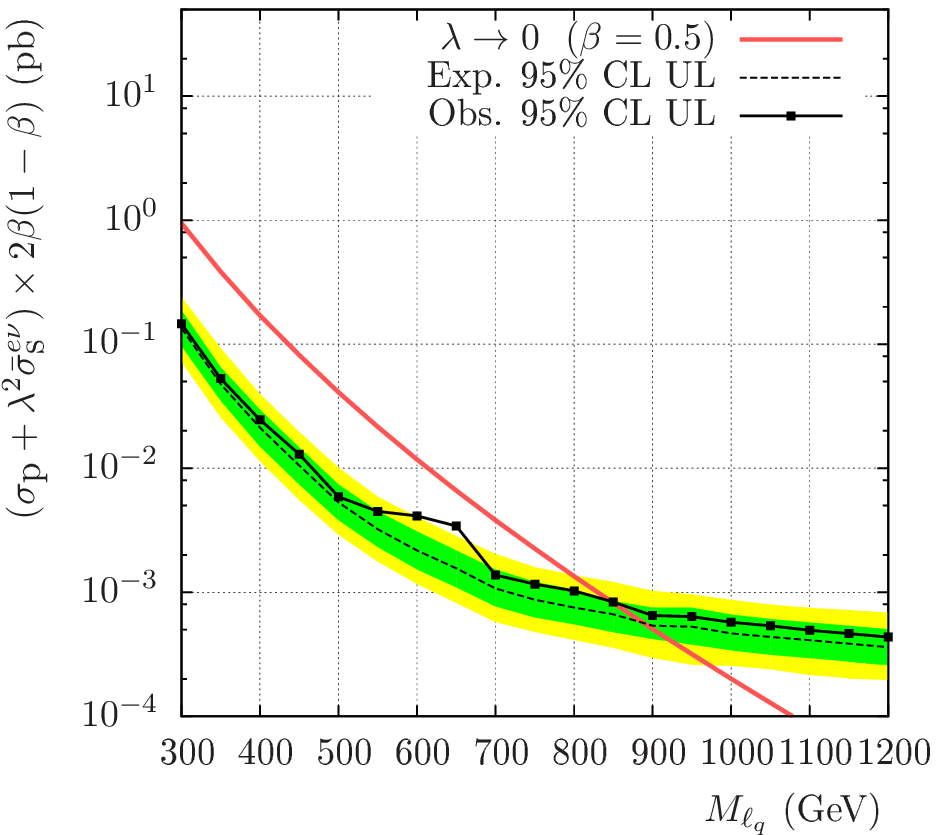}\label{fig:sig_mass_evjj_CMS}} &\\
&\subfloat[$eejj$ channel (Model A: Q$_{\rm EM}(\lq)=-1/3,5/3$)]{\includegraphics[width=\linewidth]{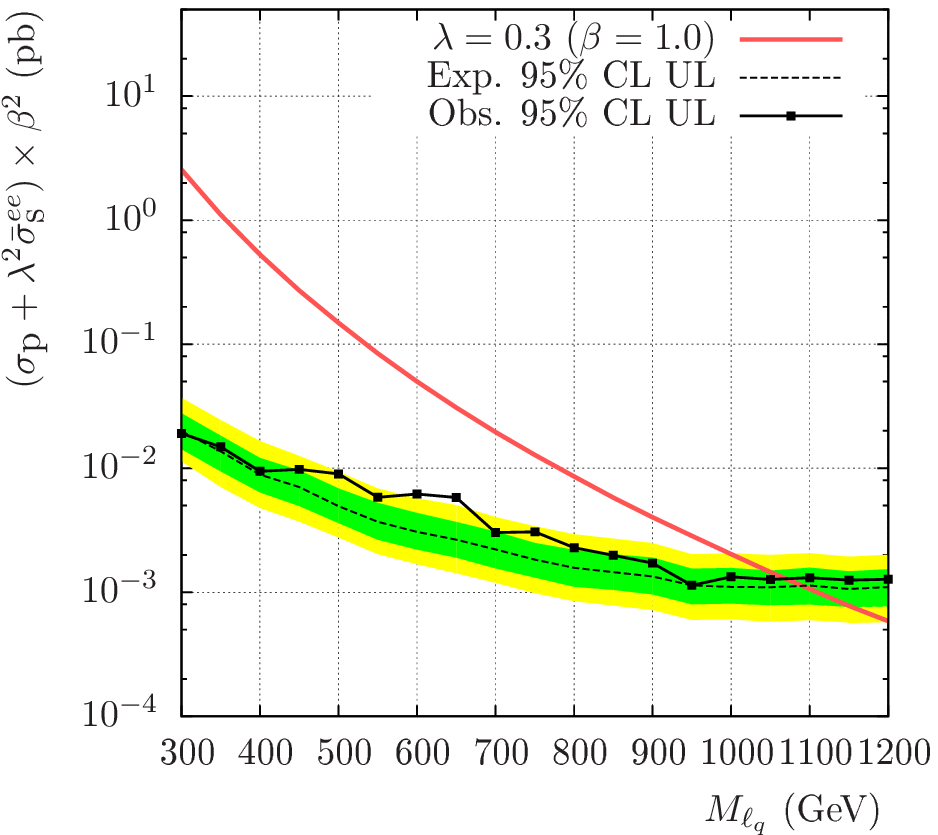}\label{fig:sig_mass_eejj_LM0_3}} 
&&
\subfloat[$e\n jj$ channel (Model A: Q$_{\rm EM}(\lq)=-1/3$))]{\includegraphics[width=\linewidth]{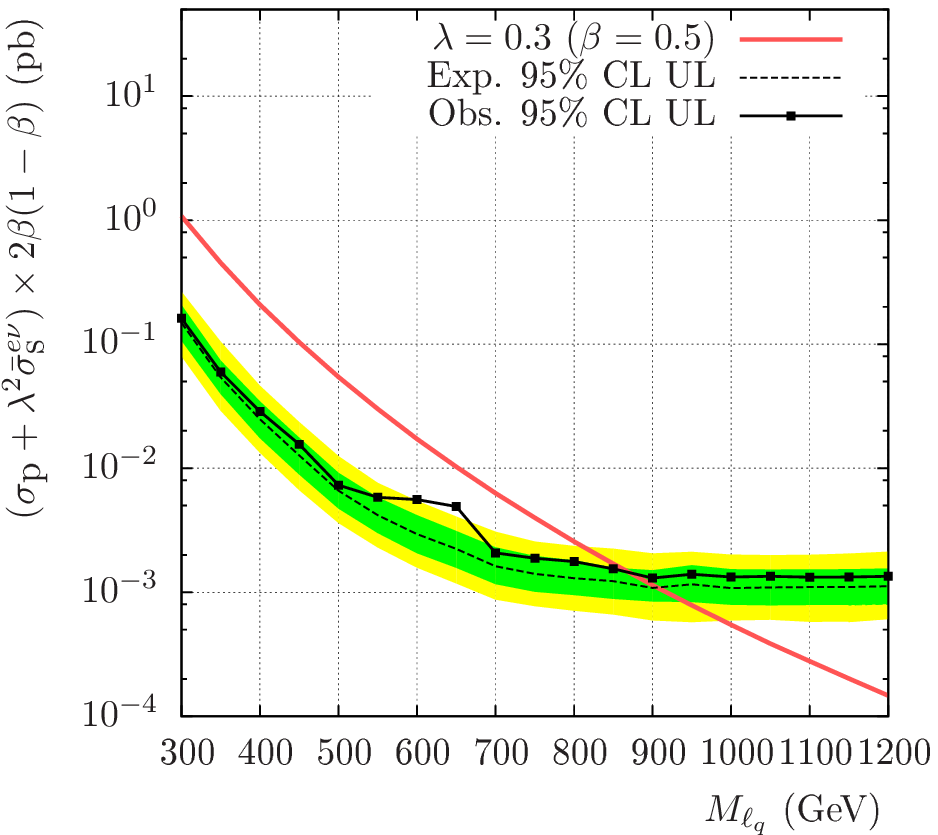}\label{fig:sig_mass_evjj_LM0_3}}&
\end{tabular}
\caption{Effect of single productions on exclusion limits. On the left panel: $eejj$ channel and on the right panel: $e\nu jj$ channel. See Eqs. \ref{eq:sg_ee_bt_dep} \& \ref{eq:sg_ev_bt_dep} for the definitions of $\bar\sg^{ee}_{\rm s}$ and $\bar\sg^{e\n}_{\rm s}$ respectively. The red solid lines are theoretical estimations. The upper-row plots ($\lm\to 0$) are same as the ones presented in Ref.~\cite{CMS:2014qpa}. For the lower-row plots $\lt(\lm=0.3,\mbox{~the value used in the CMS analyis}\rt)$, the expected and observed upper limits at 95\% CL on the LQ production cross-sections are obtained after correcting by $\mc R^{ee/e\n}\left(\lm,M_{\lq}\rt)$ (see Eqs. \ref{eq:efficiency_rescale} \& \ref{eq:rlim}).
}\label{fig:sigma_mass_exclusion}
\end{figure*}

\begin{figure*}[!t]
\begin{tabular}{rm{0.4\linewidth}m{0.08\linewidth}m{0.4\linewidth}l}
&\subfloat[Model A: Q$_{\rm EM}(\lq)=-1/3,5/3$]{\includegraphics[width=\linewidth]{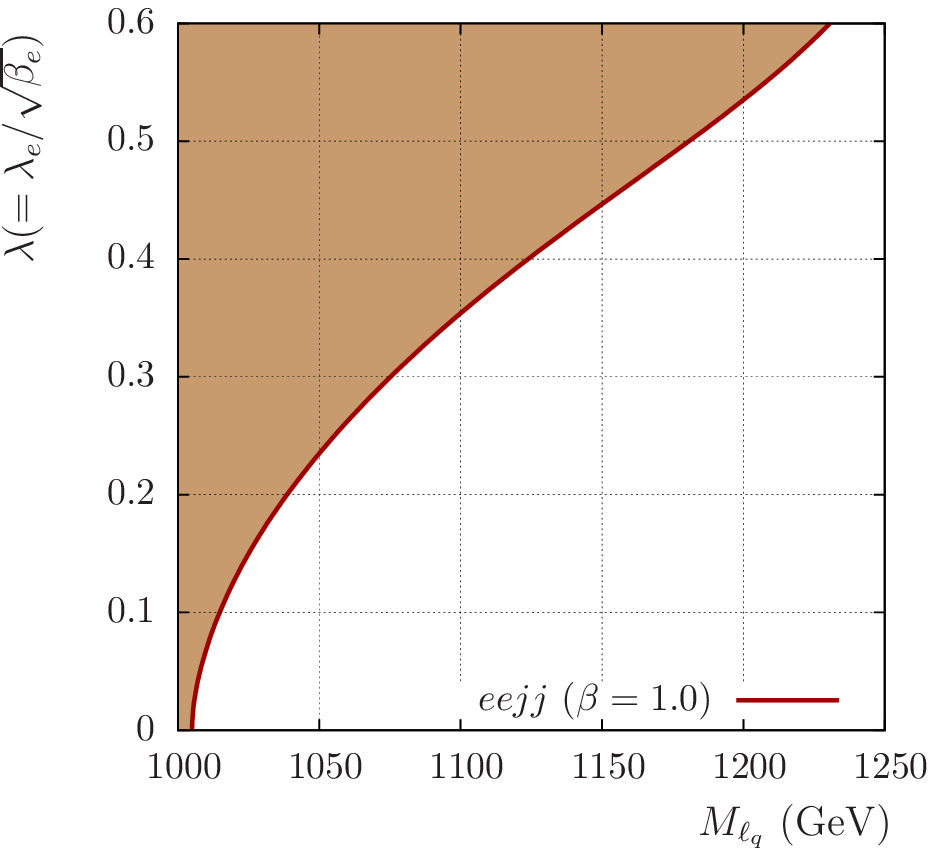}\label{fig:lm_Mexclu_bt1_0_A}}
&&
\subfloat[Model A: Q$_{\rm EM}(\lq)=-1/3$]{\includegraphics[width=\linewidth]{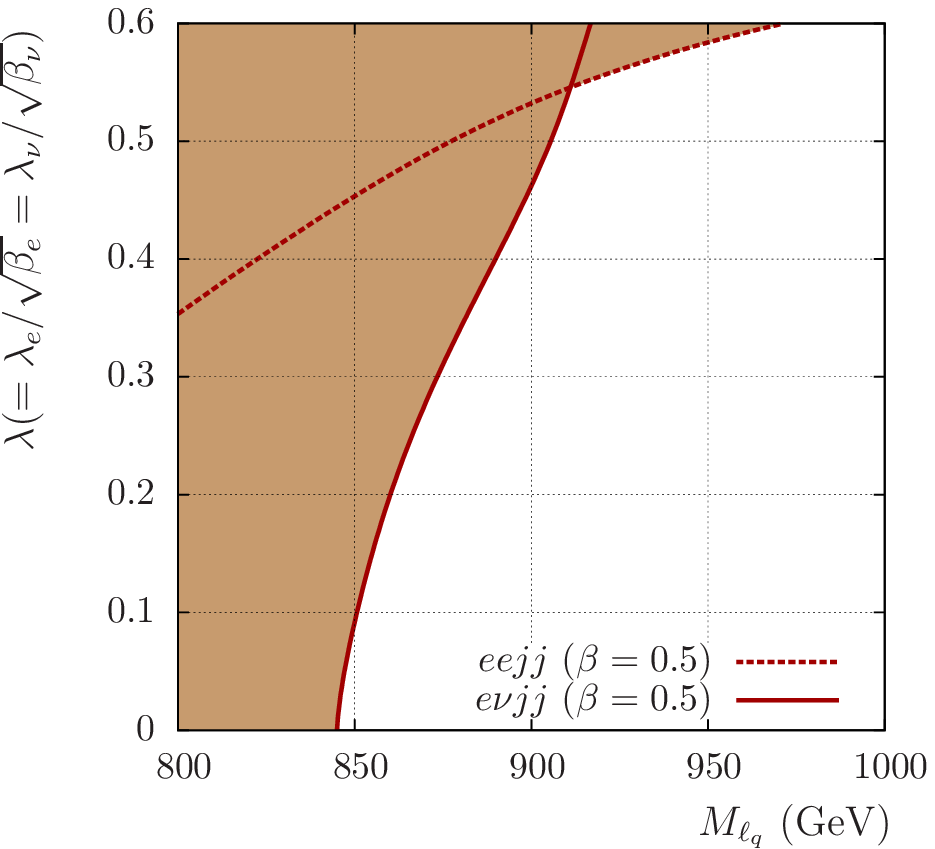}\label{fig:lm_Mexclu_bt0_5_A}}&\\
&\subfloat[Model B: Q$_{\rm EM}(\lq)=2/3,-4/3$]{\includegraphics[width=\linewidth]{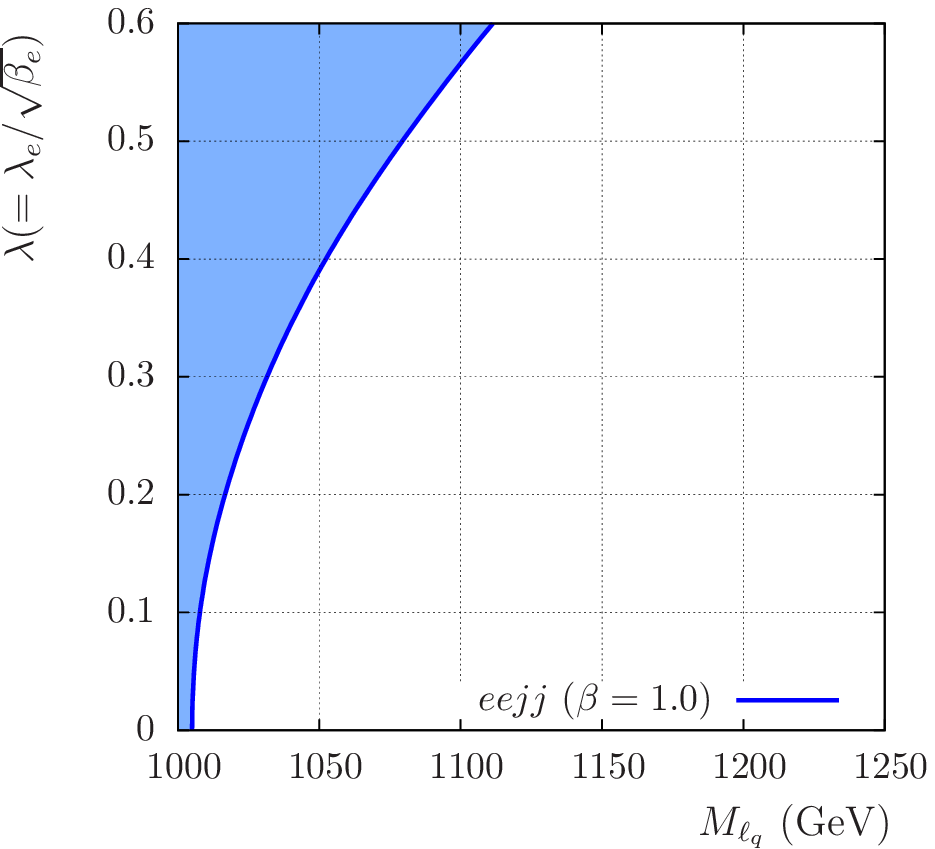}\label{fig:lm_Mexclu_bt1_0_B}}
&&
\subfloat[Model B: Q$_{\rm EM}(\lq)=2/3$]{\includegraphics[width=\linewidth]{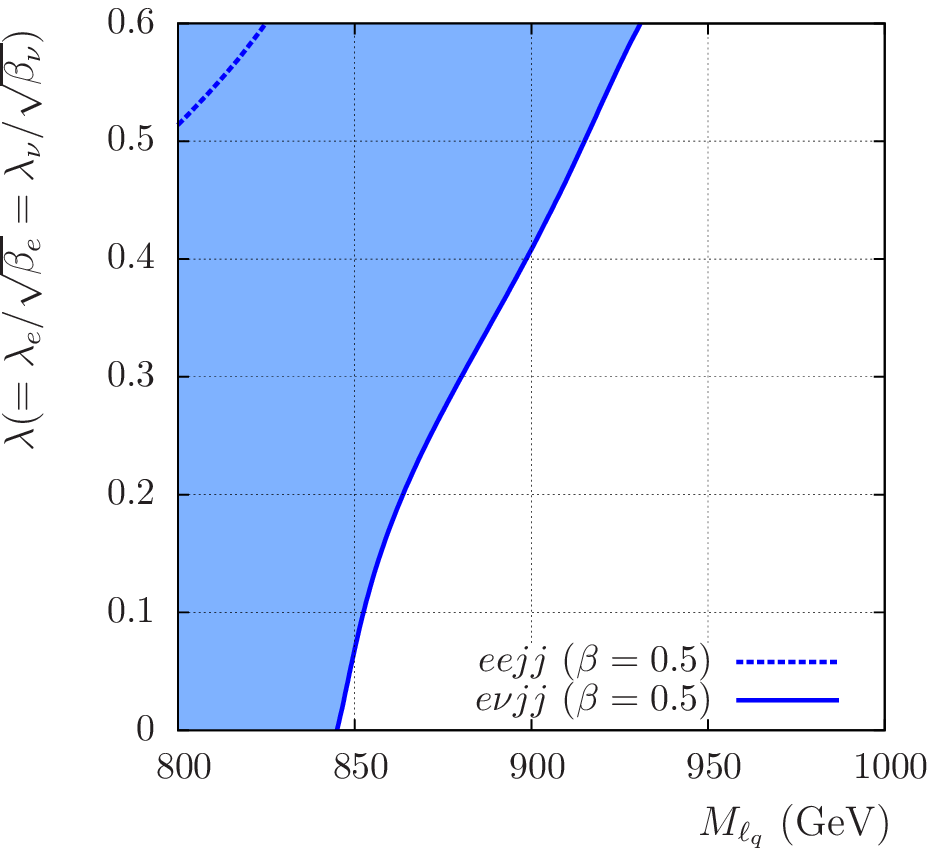}\label{fig:lm_Mexclu_bt0_5_B}}&\\
\end{tabular}
\caption{The exclusion limits (solid or dashed lines) from the observed data in $\lm-M_{\lq}$ plane for $\bt=1.0$ (left pannel) and $\bt=0.5$ (right pannel) for Models A and B. All points on the left of the solid or dashed lines (shaded regions) are excluded by the data at 95\% CL. When $\bt=0.5$, the strongest limits come from the $e\n jj$ data for both models except in Model A the $eejj$ data gives the strongest limit for $\lm\gtrsim 0.55$ ($M_{\lq} \gtrsim 910$ GeV). }\label{fig:lm_mass_exclusion}
\end{figure*}

\subsection{The $e\nu jj$ channel}\label{sec:evjj}

For the $e\nu jj$ channel we set $\bt=0.5$ to generate the signal events with a similar ME$\oplus$PS computation (the argument for computing this manner remains same as before),
\be
\left. \begin{array}{lclclll}
pp 	&\to		&(\lq~e	&+	&\lq~\nu)		&\to		& ej\n\ubr{-3.5}\lbr{-1}\,,\\
pp  &\to		&(\lq~ej	&+	&\lq~\nu j)	&\to		& ej\n\ubr{-3.5}\lbr{-1}\ j\,,\label{eq:matching_ev}\\
pp 	&\to		&(\lq~ejj&+	&\lq~\nu jj)	&\to		& ej\n\ubr{-3.5}\lbr{-1}\ jj\,.
\end{array}\right\}
\ee
Like the $eejj$ cross-section, here too the $\lm$ and $\bt$ dependence of the inclusive single production cross-section can be made explicit as,
\ba
\sg^{e\nu}_{\rm s}\lt(\bt,\lm,M_{\lq}\rt)\hspace{-2cm}&&\nn\\
 &=&  \lm_e^2\ \bt_\n\ \bar\sg^{e\nu}_{\rm s1}\left(M_{\lq}\right) 
+\lm_\nu^2\ \bt_e\ \bar\sg^{e\nu}_{\rm s2}\left(M_{\lq}\right)
+ \cdots\nn\\
 &=&  \lm^2\ \bt_e\bt_\n\ \bar\sg^{e\nu}_{\rm s1}\left(M_{\lq}\right) 
+\lm^2\ \bt_e\bt_\n\ \bar\sg^{e\nu}_{\rm s2}\left(M_{\lq}\right)
+ \cdots\nn\\
&\overset{\rm def.}{=} & \lm^2\ 2\bt\lt(1-\bt\rt)\ \bar\sg^{e\nu}_{\rm s}\left(M_{\lq}\right) + \cdots,\label{eq:sg_ev_bt_dep}
\ea
where, like before, the neglected terms are at least $\mc O(\lm^4)$ and
\ba
 \bar\sg^{e\nu}_{\rm s}\left(M_{\lq}\right) = \frac12\lt(\bar\sg^{e\nu}_{\rm s1}\left(M_{\lq}\right) + \bar\sg^{e\nu}_{\rm s2}\left(M_{\lq}\right)\rt).
\ea
Again we see that with the chosen parametrization, the $\bt$ dependence of the single production cross-section becomes the same as the pair production.

\begin{figure*}[!t]
\begin{tabular}{rm{0.4\linewidth}m{0.08\linewidth}m{0.4\linewidth}l}
&
\subfloat[$eejj$ channel (Model A: Q$_{\rm EM}(\lq)=-1/3,5/3$)]{\includegraphics[width=\linewidth]{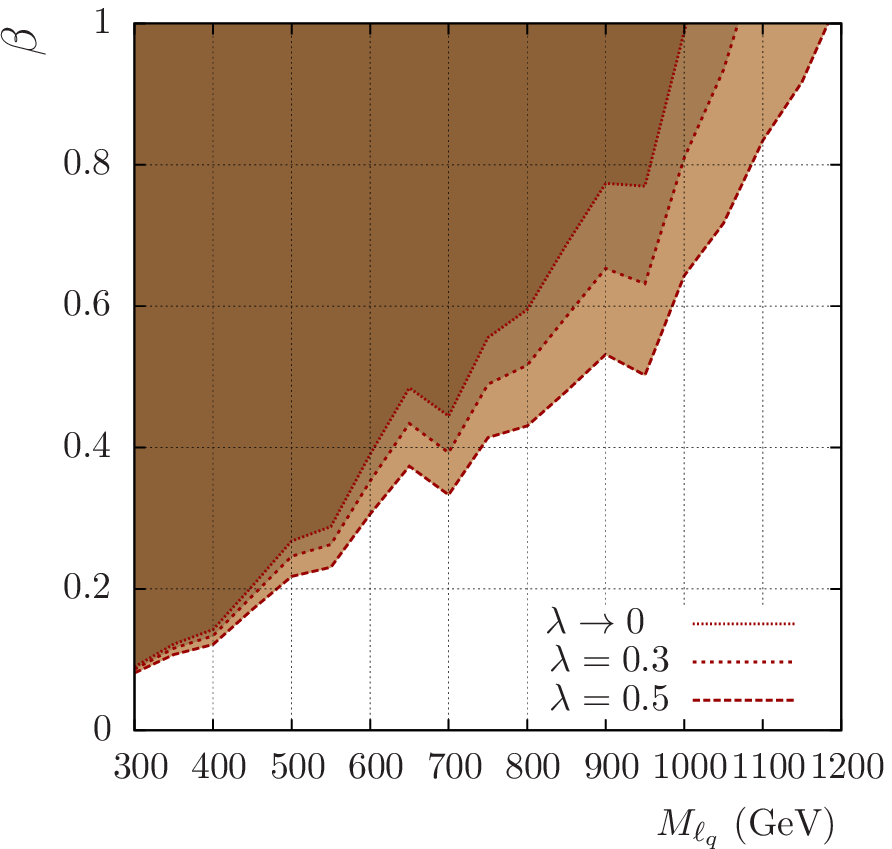}\label{fig:bt_mass_eejj_obs_A}} 
&&
\subfloat[$e\n jj$ channel (Model A: Q$_{\rm EM}(\lq)=-1/3$)]{\includegraphics[width=\linewidth]{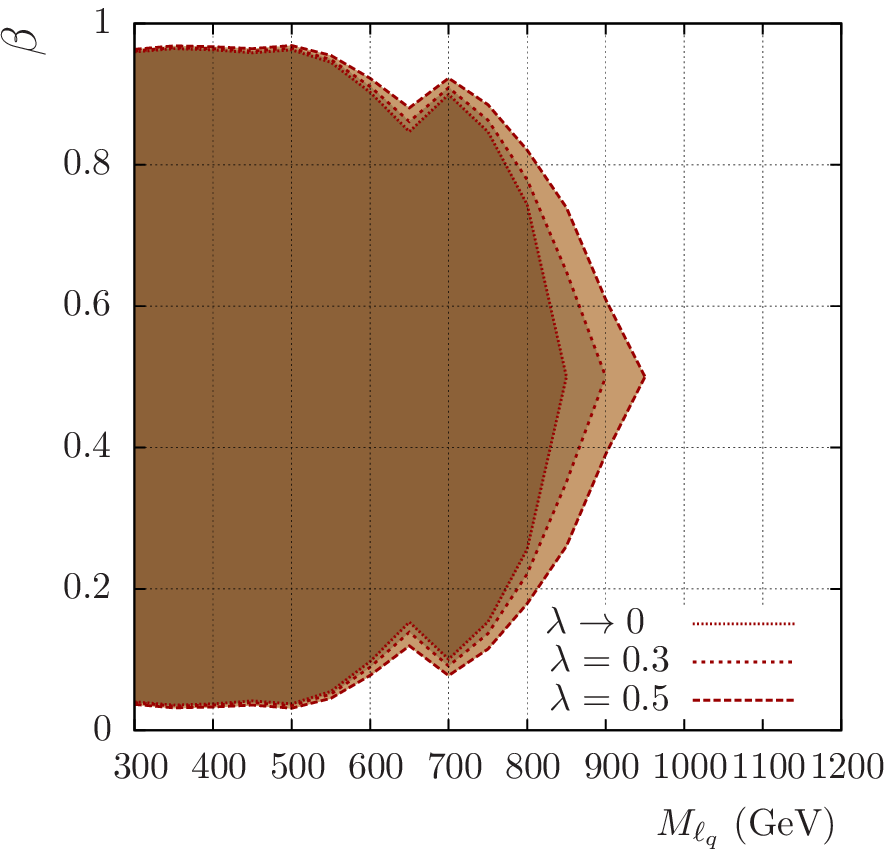}\label{fig:bt_mass_evjj_obs_A}}&\\
&
\subfloat[$eejj$ channel (Model B: Q$_{\rm EM}(\lq)=2/3,-4/3$)]{\includegraphics[width=\linewidth]{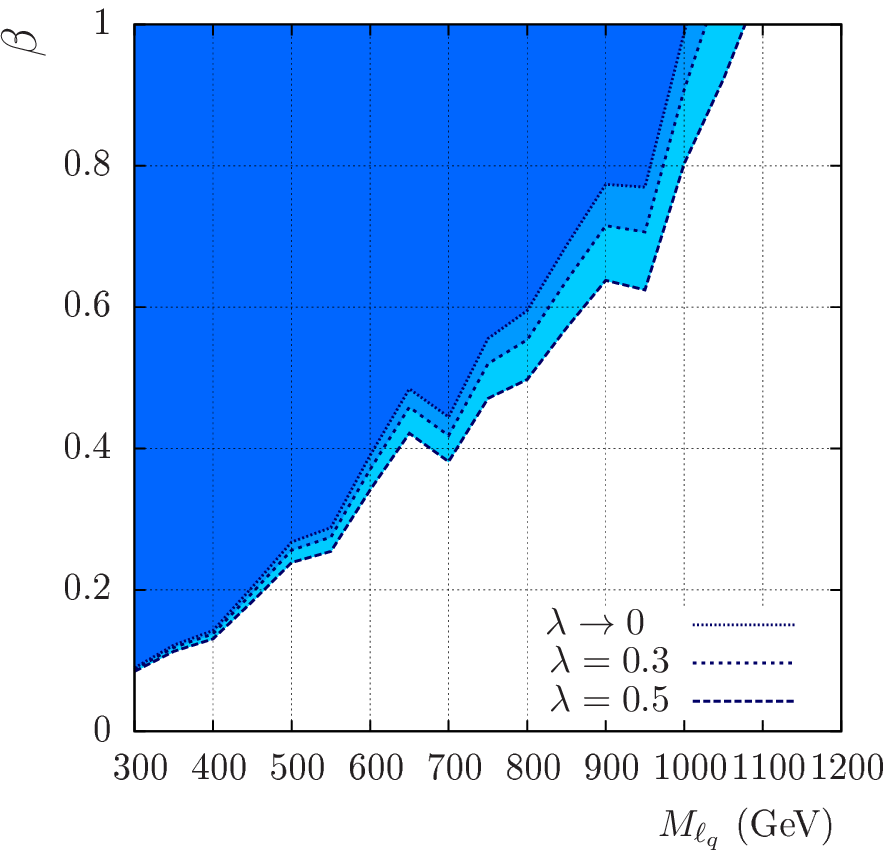}\label{fig:bt_mass_eejj_obs_B}} 
&&
\subfloat[$e\n jj$ channel (Model B: Q$_{\rm EM}(\lq)=2/3$)]{\includegraphics[width=\linewidth]{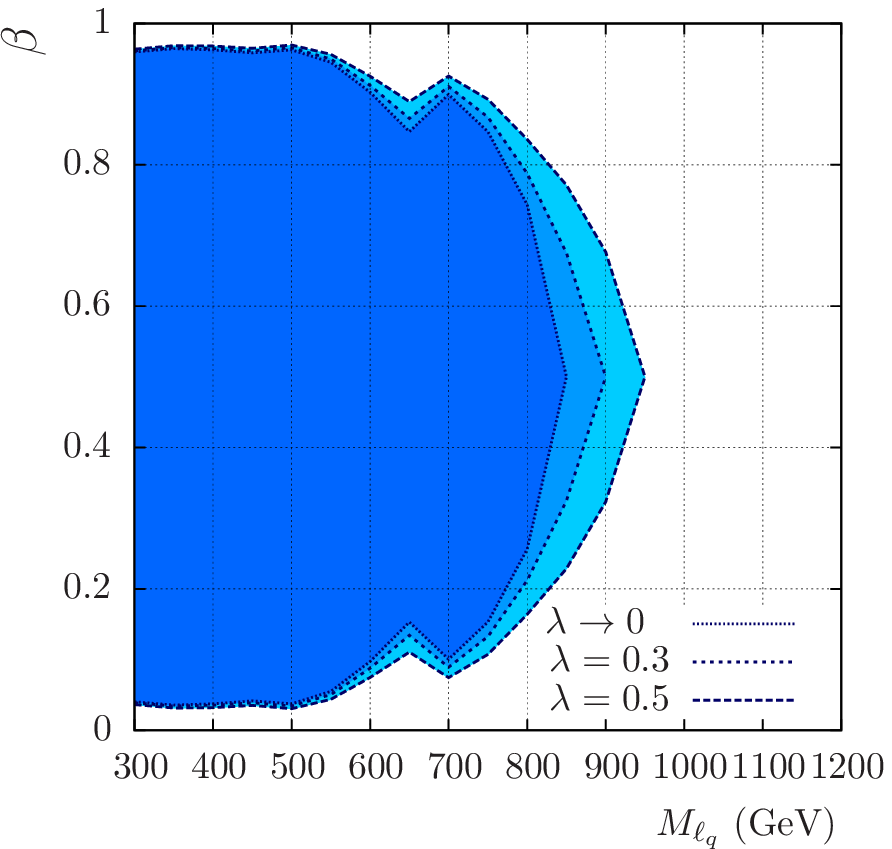}\label{fig:bt_mass_evjj_obs_B}}&
\end{tabular}
\caption{Regions excluded at 95\% CL by the observed data in the $\bt-\lm$ plane for different values of $\lm$ in Models A and B. The plots for $\lm\to0$ are obtained from pair production only and hence are same as the ones presented by CMS \cite{CMS:2014qpa}.}\label{fig:bt_mass_exclusion}
\end{figure*}

\begin{figure*}[!t]
\begin{tabular}{rm{0.4\linewidth}m{0.08\linewidth}m{0.4\linewidth}l}
&
\subfloat[$eejj$ channel (Model A: Q$_{\rm EM}(\lq)=-1/3,5/3$)]{\includegraphics[width=\linewidth]{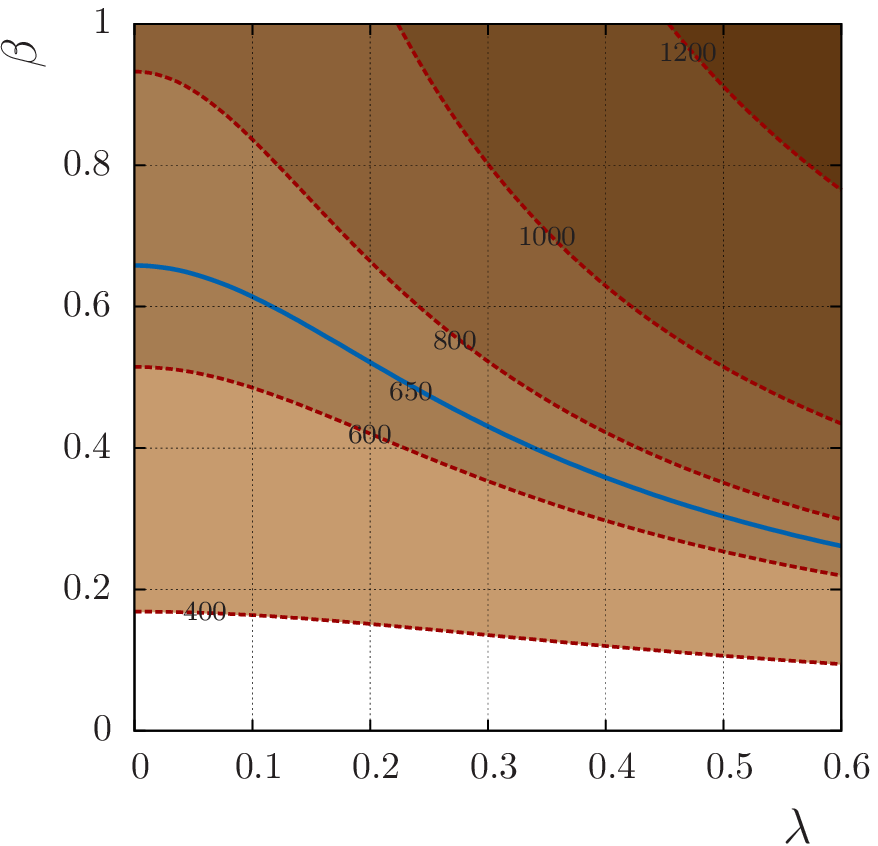}\label{fig:btlam_eejj_obs_A}} 
&&
\subfloat[$e\n jj$ channel (Model A: Q$_{\rm EM}(\lq)=-1/3$)]{\includegraphics[width=\linewidth]{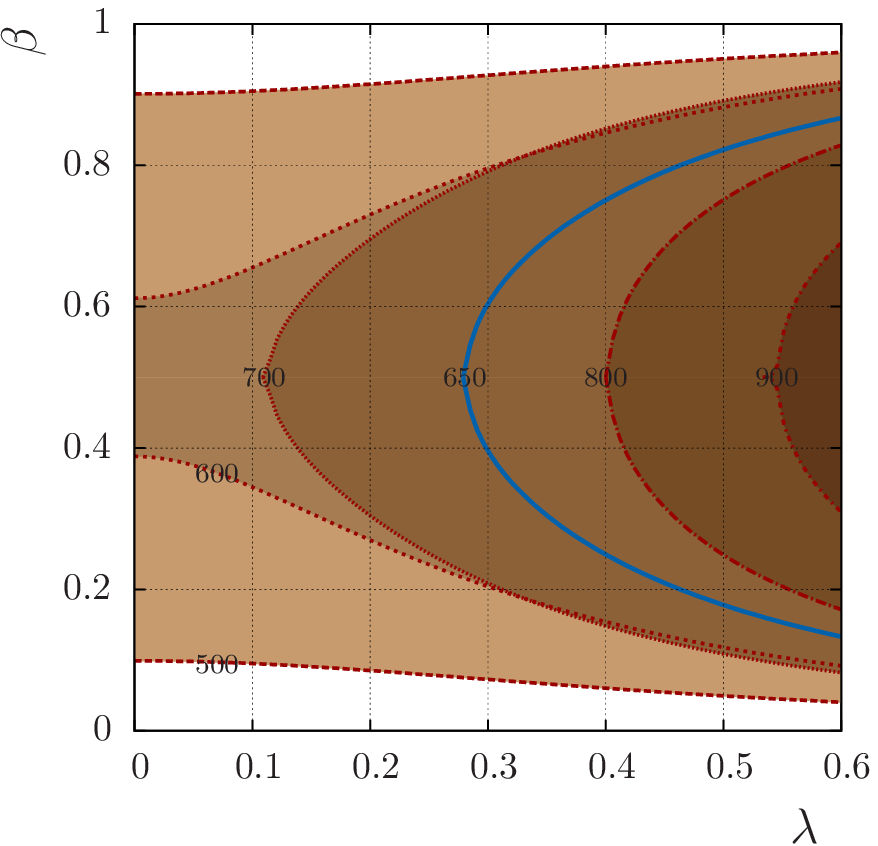}\label{fig:btlam_evjj_obs_A}}&\\
&
\subfloat[$eejj$ channel (Model B: Q$_{\rm EM}(\lq)=2/3,-4/3$)]{\includegraphics[width=\linewidth]{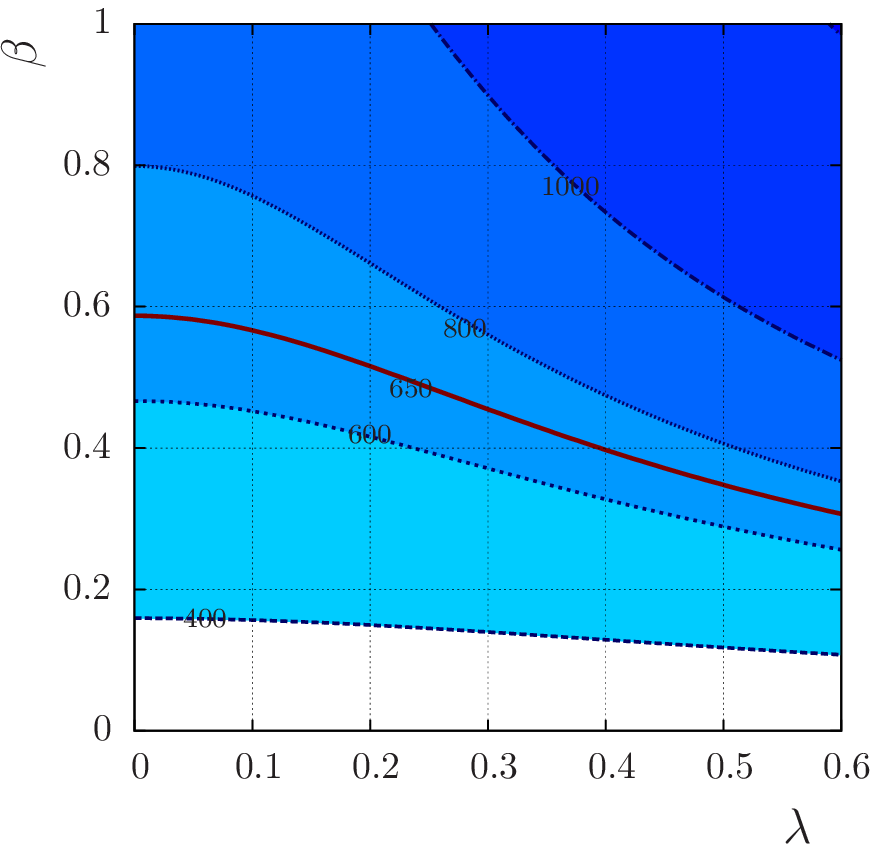}\label{fig:btlam_eejj_obs_B}} 
&&
\subfloat[$e\n jj$ channel (Model B: Q$_{\rm EM}(\lq)=2/3$)]{\includegraphics[width=\linewidth]{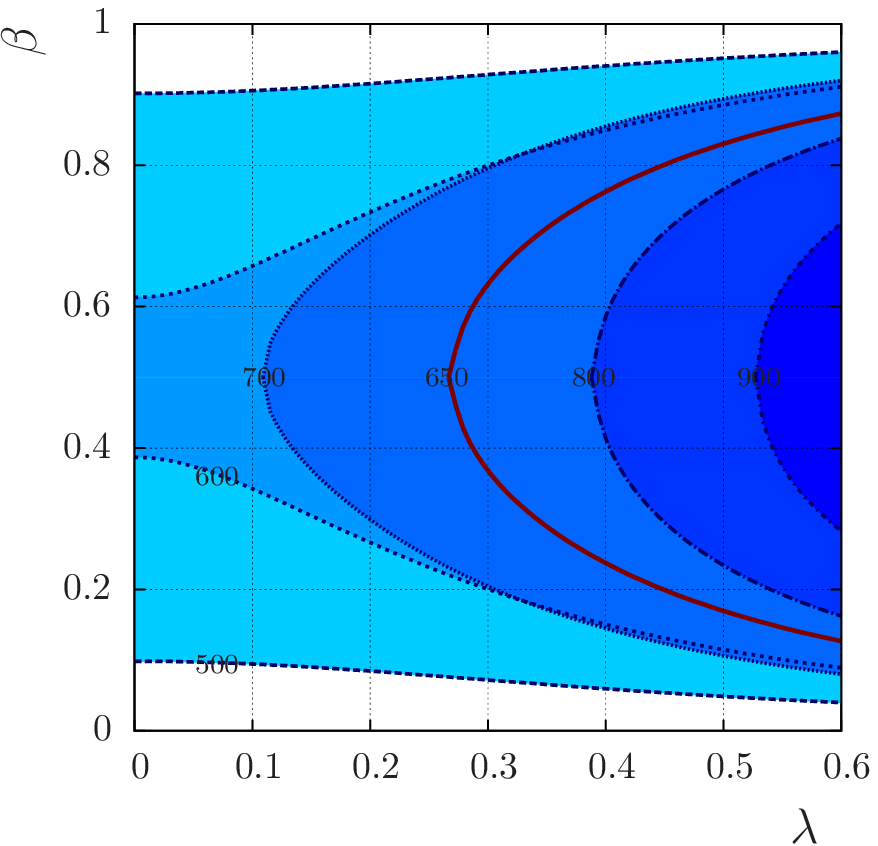}\label{fig:btlam_evjj_obs_B}}&
\end{tabular}
\caption{Regions excluded at 95\% CL by the observed data in the $\bt-\lm$ plane for different values of $M_{\lq}$ in Models A and B. On the left panel: $eejj$ channel and on the right panel: $e\nu jj$ channel. The exclusion limits for fixed $M_{\lq}$ values are shown as boundaries between differently shaded regions (except the $M_{lq}=650$ GeV lines which are drawn explicitly). For the $eejj$ channel, the region above any fixed mass line is excluded whereas for $e\n jj$ channel the region on the right of any fixed mass line is excluded.}\label{fig:bt_lam_exclusion}
\end{figure*}

\section{Results: rescaled exclusion limits }\label{sec:results}
In Table \ref{tab:sigma} we show the theoretical cross-sections for pair production at the LO and inclusive single productions (in Model A) for $eejj$ and $e\n jj$ channels with $M_{\lq}=$ 550, 650, 750 GeV and $\lm=$ 0.3, 0.5. As expected, the pair production is almost independent of $\lm$ whereas the single production cross-section increases as $\lm^2$. 
In Table \ref{tab:evntscms} we show the number of events that pass the final selection criteria for pair production ($\mc N_{\rm p}$) (from Ref. \cite{CMS:2014qpa}, obtained with $\sg^{\rm NLO}_{\rm p}$\cite{Kramer:2004df}) and inclusive single production ($\mc N_{\rm s}$) for both $eejj$ or $e\nu jj$ channels. In the same table we also show the selection-cut efficiency for the single production, $\ep_{\rm s}$ -- the fraction of events that survives the selection criteria for a particular $M_{\lq}$,
\ba
\ep_{\rm s}^{\rm X}(M_{\lq}) =  \frac{\mc N^{\rm X}_{\rm s} }{ \sg_{\rm s}^{\rm X}\times \mc L}\;, \label{eq:efficiency_single}
\ea
where $\mc L=19.6$ fb$^{-1}$ is the integrated luminosity for which $\mc N_{\rm s}$ is computed and $X=\{ee$ or $e\n\}$.
 Note that $\ep^{\rm X}_{\rm s}(M_{\lq})$ does not depend on any overall factors in $\sg_{\rm s}$  like $\lm^2, \bt_e^2$ (for $eejj$ channel) or $\bt_e\bt_\n$ (for $e\n jj$ channel) etc. The numbers show that even for $\lm$ as small as 0.3 a good number of events survive the selection cuts optimized for the pair production. Moreover the ratio of $\mc N_{\rm s}/\mc N_{\rm p}$ increases with increasing $M_{\lq}$, i.e., with increasing mass, single productions contribute more and more compared to the pair production.

In Fig. \ref{fig:sigma_mass_exclusion}, we show how the presence of single productions affects mass ELs. In the vertical axes of these plots we have now plotted 
\ba
\bt^2\sg_{\rm p}+\sg_{\rm s}^{ee}\lt(\bt,\lm\rt) = \bt^2\lt(\sg_{\rm p}+\lm^2\bar\sg_{\rm s}^{ee}\rt)
\ea 
instead of $\bt^2\sg_{\rm p}$ for the $eejj$ channel and 
\ba 
2\bt\lt(1-\bt\rt)\sg_{\rm p}+\sg_{\rm s}^{e\n}\lt(\bt,\lm\rt) = 2\bt\lt(1-\bt\rt)\lt(\sg_{\rm p}+\lm^2\bar\sg_{\rm s}^{e\n}\rt)
\ea
instead of $2\bt\lt(1-\bt\rt)\sg_{\rm p}$ for the $e\n jj$ channel where $\bar\sg^{ee}_{\rm s}$ and $\bar\sg^{e\n}_{\rm s}$ are defined in Eqs. \ref{eq:sg_ee_bt_dep} \& \ref{eq:sg_ev_bt_dep} respectively. When $\lm\to 0$, single productions vanish, hence the upper row plots are identical to the ones presented by CMS (we avoid writing $\lm=0$ as in that case both the decay channels would vanish). For the plots with $\lm=0.3$, we have rescaled the expected and observed 95\% CL  upper limits on the LQ pair production cross-sections obtained by CMS by multiplying them with a factor
\ba
 \mc R^{\rm X}\lt(\lm,M_{\lq}\rt) = \frac{\ep_{\rm p}^{\rm X}\lt(M_{\lq}\rt)}{\ep_{\rm p+s}^{\rm X}\lt(\lm,M_{\lq}\rt)}\,,\label{eq:efficiency_rescale}
\ea
where $\ep_{\rm p}$ is the selection-cut efficiency for pair production defined in the same manner as $\ep_{\rm s}$ in Eq. \ref{eq:efficiency_single},
\ba
\ep_{\rm p}^{\rm X}(M_{\lq}) =  \frac{\mc N^{\rm X}_{\rm p} }{\lt(f^{\rm X}\lt(\bt\rt)\,\sg_{\rm p}^{\rm NLO}\rt)\times\mc L}\;, \label{eq:efficiency_pair}
\ea
with
\ba
f^{\rm X}\lt(\bt\rt) =&\bt^2&\mbox{~for~} {\rm X}=ee\;, \nn\\
=& 2\bt\lt(1-\bt\rt)&\mbox{~for~} {\rm X}=e\n\;
\ea
 and $\ep_{\rm p+s}\lt(\lm,M_{\lq}\rt)$ is defined as follows,
\ba 
\ep_{\rm p+s}^{\rm X}\lt(\lm,M_{\lq}\rt)\hspace{-2.25cm}&&\nn\\
&=&\frac{f^{\rm X}\lt(\bt\rt)\sg_{\rm p}\lt(M_{\lq}\rt)\ep_{\rm p}^{\rm X}\lt(M_{\lq}\rt) + \sg_{\rm s}^{\rm X}\lt(\bt,\lm, M_{\lq}\rt)\ep_{\rm s}^{\rm X}\lt(M_{\lq}\rt)}{f^{\rm X}\lt(\bt\rt)\sg_{\rm p}\lt(M_{\lq}\rt) + \sg_{\rm s}^{\rm X}\lt(\bt,\lm, M_{\lq}\rt)} \nn \\ 
&=&\frac{\sg_{\rm p}\lt(M_{\lq}\rt)\ep_{\rm p}^{\rm X}\lt(M_{\lq}\rt) + \lm^2\bar\sg_{\rm s}^{\rm X}\lt(M_{\lq}\rt)\ep_{\rm s}^{\rm X}\lt(M_{\lq}\rt)}{\sg_{\rm p}\lt(M_{\lq}\rt) + \lm^2\bar\sg_{\rm s}^{\rm X}\lt(M_{\lq}\rt)}\;.
\ea
When $\lm\to 0$, $\sg_{\rm s}^{\rm X}\lt(\bt,\lm, M_{\lq}\rt)$ vanishes and hence
\ba
\lim_{\lm\to 0}\mc R^{\rm X}\lt(\lm,M_{\lq}\rt)=1\;.\label{eq:rlim}
\ea 
For our analysis, we have used the numbers given in the CMS analysis to compute $\ep_{\rm p}$.

We can see how the 95\% CL ELs change with $\lm$ from Fig. \ref{fig:lm_mass_exclusion} which shows the observed ELs in the $\lm-M_{\lq}$ plane for $\bt=1.0$ and $\bt=0.5$ and both Models A and B.  These plots can also be used to put upper limits on $\lm$ (or equivalently on $\lm_e$ and $\lm_\n$ for any $\bt$) for certain values of $M_{\lq}$.
Once single productions are considered, the limit coming from the $eejj$ data for $\lm=$ 0.3 (0.6) on the mass of the LQs with Q$_{\rm EM}=-1/3,5/3$  goes up to about 1070 (1230) GeV from 1005 GeV for $\bt=1.0$. However, for the LQs with Q$_{\rm EM}=2/3,-4/3$, the corresponding limit reaches only about 1030 (1110) GeV.  
When $\bt=0.5$, the most stringent limits come from the $e\n jj$ data for both models except in Model A where the $eejj$ data gives the strongest limit for $\lm\gtrsim 0.55$ ($M_{\lq} \gtrsim 910$ GeV). For a LQ with Q$_{\rm EM}=-1/3$, the mass EL improves from 845 GeV to about 870 (970) GeV for $\lm=0.3$ (0.6). For the LQs  with Q$_{\rm EM}=2/3$ the same limit improves to about 880 (930) GeV.
The excluded regions in the $\bt-M_{\lq}$ plane are shown in Fig. \ref{fig:bt_mass_exclusion} for $\lm\approx 0$, 0.3 and 0.5 respectively. The plots show that for any fixed $\bt$, the ELs increase with increasing $\lm$.
In Fig. \ref{fig:bt_lam_exclusion}, we show how the observed ELs vary with $\bt$ in the $\bt-\lm$ plane for the two channels in both Models A and B. Notice, that in the $e\n jj$ channel plots  (Figs. \ref{fig:btlam_evjj_obs_A} \& \ref{fig:btlam_evjj_obs_B}), the EL curves for $M_{\lq}=650$ GeV lie among the $M_{\lq}=700$ and 800 GeV curves. This happens because of the observed excess at 650 GeV in the data.

Inclusion of single productions in the analysis can somewhat improve the $\chi^2$-fit of the observed data in $eejj$ and $e\n jj$ channels separately. We use the following $\chi^2$ function,
\ba
\chi^2\lt(\bt,\lm\rt) = \sum_{i}\lt(\frac{N_{\rm LQ}^i\lt(\bt,\lm\rt) + N_{\rm BG}^i - N^{i}_{\rm Data}}{\Dl N_{\rm BG}^i}\rt)^2,
\ea 
where $N_{\rm LQ}$, $N_{\rm BG}$ and $N_{\rm Data}$ are the number of LQ signal events (MC), SM background events \cite{CMS:2014qpa} (MC) and observed events (data), respectively for the $i^{\rm th}$ benchmark LQ mass taken by CMS (varied from 300 GeV to 1200 GeV in steps of 50 GeV). The errors, $\Dl N_{\rm BG}^i$ are obtained by adding the systematic and statistical uncertainties of the total background in quadrature. 
In Model A with $M_{\lq}=650$ GeV and $\lm=0.6$, as we move from `only pair' to `pair+single', the best fitted $\bt$ changes from 0.19 (0.044) to 0.15 (0.025) in the the $eejj$ ($e\nu jj$) channel.  The corresponding decrease in the minimum of $\chi^2$, ($\Dl\chi^2_{min}$) is about 18\% (15\%). Similarly, for Model B, the best fitted $\bt$ changes from 0.19 (0.044) to 0.17 (0.022) in the the $eejj$ ($e\nu jj$) channel and the corresponding $\Dl\chi^2_{min}$ is about 13\% (11\%).
However, when we combine the two channels $\Dl\chi^2_{min}$ becomes insignificant in both the models. 

\begin{figure}[!t]
\begin{center}
\includegraphics[width=\linewidth]{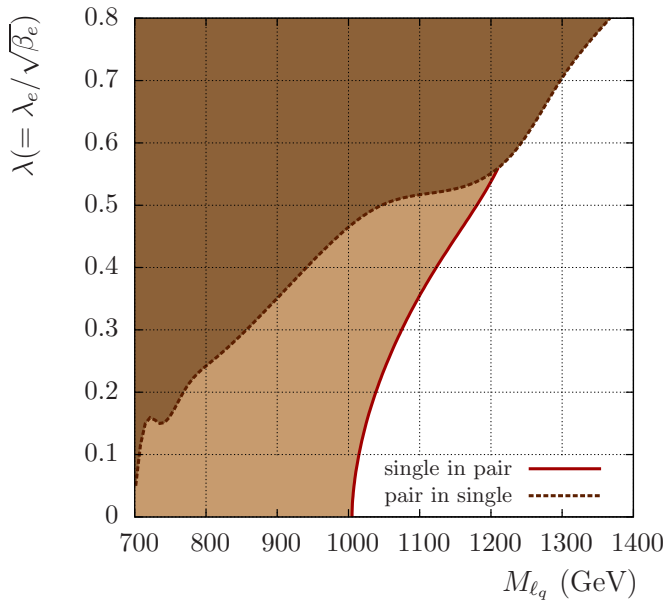}
\caption{Comparison of 95\% ELs obtained by including the contribution of single productions in the pair production search data \cite{CMS:2014qpa} (single in pair) and that of the pair production in the single production search data \cite{CMS:2015kda} (pair in single). Here we have considered Model A to simulate the single production events and set $\bt=\bt_e=1$. \label{fig:lm_Mexclu_CMSsing}}
\end{center}
\end{figure}
Before we move on, we make a small digression here. Very recently, CMS has presented a new dedicated search for single production of scalar LQs with a final state of two electrons and one jet (for the first generation) \cite{CMS:2015kda}.
If we compare roughly, for Model A with $\lm=0.6$ and $\bt=1$, our rescaled EL from the pair production data on $M_{\lq}$ (1230 GeV) is competitive to the one obtained (1260 GeV) there. For $\lm=0.4$ and $\bt=1$, the rescaled limit actually gives a better estimation of the lower bound on $M_{\lq}$ than the single production, because the former one includes the contribution from pair production. Now, just as we have included the contribution of single productions in the pair production search, we can apply the same logic and consider the contamination from the pair production in this dedicated search for single production by similar rescaling. In fact, in this case one can not avoid the pair production completely by tuning $\lm$, since it is almost independent of this parameter. In Fig. \ref{fig:lm_Mexclu_CMSsing} we compare the limits thus obtained from these two cases in the  $\lm-M_{\lq}$ plane. The EL obtained by including the single production contribution in the pair production search (i.e. same as in Fig. \ref{fig:lm_Mexclu_bt1_0_A}) is shown by the solid line and the one obtained by doing vice versa is shown by the dashed line. For $\lm\lesssim 0.55$, the pair production search data gives the better limit on $M_{\lq}$. However, with increasing $\lm$ and (excluded) $M_{\lq}$, single production contribution increases and pair production contribution goes down. For $\lm \gtrsim 0.55$ and $ M_{\lq}  \gtrsim$ 1200 GeV, the single productions take over the pair production to determine the EL. For such large $\lm$ and $M_{\lq}$, the contribution of the pair production becomes negligible and hence, in this range, both limits converge to the one obtained by considering single productions only \cite{CMS:2015kda}.

\section{Discussions \& Conclusions}\label{sec:conclusion}
One must be careful while interpreting our results since they are obtained by simple rescaling, not a full statistical analysis. Nonetheless, they certainly show that within the probed range of $M_{\lq}$, systematic inclusion of the single productions in the signal simulation of the pair production search improves the ELs. For $\lm=\lm_{\rm QED}=0.3$ (the value used for MC simulations) the CMS analysis in Ref.~\cite{CMS:2014qpa} underestimates the mass ELs. This is just a manifestation of the fact that the reach of the LHC (to probe new particles) can increase if multiple production channels are systematically combined (we have already demonstrated this while estimating the LHC discovery reach for leptogluons which also decay to $\ell j$ pairs like LQs \cite{Mandal:2012rx}, support of this argument can also be found in Ref. \cite{Belyaev:2005ew}). While it can be argued that the pair production gives the most conservative estimation of ELs on mass of LQs (or any colored particles), but, as we have seen, it can be misleading about other parameters like the branching fraction, $\bt$. Moreover, by ignoring single productions completely, one also ignores the fact that the pair production data is not only excluding masses of new particles but also putting limits on new couplings. For LQs, the limits on $\lm$ thus obtained are competitive with the ones obtained from HERA observations \cite{Collaboration:2011qaa,Abramowicz:2012tg}. Our na\"ive estimates indicate that compared to the limits from HERA, the CMS data already provides better limits (both in terms of lower limits on $M_{\lq}$ and upper limits on $\lm$) for the entire parameter range covered in Fig. \ref{fig:lm_mass_exclusion}. 

The results also support our argument that for any fixed $\lm$, the ELs depart more and more from the pair production ones as $M_{\lq}$ increases.   Hence, if single productions must be ignored, it also has to be made certain that within the whole mass range probed, the single productions remain small enough and the choice of $\lm$ should reflect this. Alternatively, the selection-cuts could be optimized in such a way that they disfavor single productions strongly. But, rather than making any assumptions like this, our suggestion is to include single productions in the signal simulations and allow $\lm$ to vary as much as the existing experimental limits allow. Moreover, if the selection cuts could also be optimized so that they do \emph{not} disfavor the single productions unnecessarily, then it may be possible to impinge  more on the presently allowed parameter space due to increased sensitivity to $\lm$ (also see e.g. Refs. \cite{Belyaev:2005ew,Mandal:2012rx}). Also, advanced techniques (like multivariate analysis) could tell us more about parameters other than the mass. This way, from the same experiment, we might learn more about the underlying model than a conservative mass exclusion of LQs.

So far our discussions have been mainly centred on the example of LQs. However, it is easy to see that the argument in favour of including single productions is applicable for many other BSM particle searches. Generally, if there is no symmetry that prevents model dependent production (like R-parity prevents single production of supersymmetric particles), ideally all such productions that could (potentially) contribute to their search should also be considered. At the core of our argument stands the very simple and well known statement -- \emph{ideally}, all processes that can have similar final states should be included in the signal simulations. However, \emph{in practice,} one often has to make simplifying assumptions like, if there are too many processes to produce a new particle that could contribute to its search, usually the `sub-dominant' ones are ignored, or if there are different new particles that can have similar signals, generally only one is considered at a time etc.
Here, we are simply trying to suggest that when making such simplifying assumptions one should justify them \emph{quantitatively} to the extent possible and, moreover, if technology permits one need not make any `unnecessary' assumptions. For example, in the context of LQs, for very large $M_{\lq}$ (say $\sim$ 1500 GeV) there is no experimental bound that prevents $\lm$ from being order one. Hence, while probing such masses one need not set $\lm$ to be very small. Similarly, as our estimation suggests, a dedicated search for single production of LQs at the LHC should also consider the possible contribution from pair production.

To summarize, we have shown that while searching for new colored particles at the LHC, one need not always consider model dependent (single) and (mostly) model independent (pair) productions separately.  It may be unnecessary to ignore single productions compared to the pair production  in some cases. Single productions can obviously be important if the couplings controlling them are not small -- this is why the studies that focus on single productions generally consider large couplings (see \cite{Dorsner:2014axa} for an example in the context of LQ) - but even if one assumes the couplings to be small, single productions can still contribute significantly if the mass of the particle being probed is sufficiently high. In such a case, by ignoring single productions na\"ively, one might conclude wrongly about parameters like branching fraction etc. With the example of the recent search for pair production of scalar LQs performed by CMS, we have illustrated how the contribution from single productions can be included in the analysis and how it can lead to modified limits on the LQ-mass. In addition to this, inclusion of model dependent single productions gives us new information about model couplings like the LQ-lepton-quarks couplings from the same experimental data. For this purpose, we have proposed two generic models for scalar LQs that can act as templates for a wider variety of LQs and can accommodate the experimental data easily. Applying similar logic, we have also demonstrated how the pair production could affect bounds obtained from the more recent CMS search for LQs in single production channels. Finally, we have pointed out that our arguments are not specific to the case of LQs (which we have used as an illustrative example), but they are applicable to other BSM searches too.

\begin{acknowledgements}
T.M. thanks Shilpi Jain for a helpful discussion on experimental methods. S.M. thanks Pankaj Jain for his encouragements and kind hospitality at IIT Kanpur. T.M. is partially supported by funding from the DAE, for the RECAPP, HRI. 
\end{acknowledgements}


\begin{thebibliography}{99}


\bibitem{CMS:2014qpa} 
  CMS Collaboration,
  CMS-PAS-EXO-12-041.


\bibitem{Pati:1974yy} 
  J.~C.~Pati and A.~Salam,
  Phys.\ Rev.\ D {\bf 10}, 275 (1974)
  [Erratum-ibid.\ D {\bf 11}, 703 (1975)].
  
  
\bibitem{Georgi:1974sy} 
  H.~Georgi and S.~L.~Glashow,
  Phys.\ Rev.\ Lett.\  {\bf 32}, 438 (1974).
  

\bibitem{Schrempp:1984nj} 
  B.~Schrempp and F.~Schrempp,
  Phys.\ Lett.\ B {\bf 153}, 101 (1985).

\bibitem{Kohda:2012sr} 
  M.~Kohda, H.~Sugiyama and K.~Tsumura,
  Phys.\ Lett.\ B {\bf 718}, 1436 (2013)
  [arXiv:1210.5622 [hep-ph]].
  

\bibitem{Barbier:2004ez} 
  R.~Barbier, C.~Berat, M.~Besancon, M.~Chemtob, A.~Deandrea, E.~Dudas, P.~Fayet and S.~Lavignac {\it et al.},
  Phys.\ Rept.\  {\bf 420}, 1 (2005)
  [hep-ph/0406039].
  
\bibitem{Aad:2011ch} 
  G.~Aad {\it et al.}  [ATLAS Collaboration],
  Phys.\ Lett.\ B {\bf 709}, 158 (2012)
  [Erratum-ibid.\ B {\bf 711}, 442 (2012)]
  [arXiv:1112.4828 [hep-ex]].
  
\bibitem{ATLAS:2012aq} 
  G.~Aad {\it et al.}  [ATLAS Collaboration],
  Eur.\ Phys.\ J.\ C {\bf 72}, 2151 (2012)
  [arXiv:1203.3172 [hep-ex]].

\bibitem{ATLAS:2013oea} 
  G.~Aad {\it et al.}  [ATLAS Collaboration],
  JHEP {\bf 1306}, 033 (2013)
  [arXiv:1303.0526 [hep-ex]].
  
\bibitem{Chatrchyan:2012vza} 
  S.~Chatrchyan {\it et al.}  [CMS Collaboration],
  Phys.\ Rev.\ D {\bf 86}, 052013 (2012)
  [arXiv:1207.5406 [hep-ex]].
  
\bibitem{CMS:zva} 
  CMS Collaboration,
  CMS-PAS-EXO-12-042.
  
\bibitem{Khachatryan:2014ura} 
  V.~Khachatryan {\it et al.}  [CMS Collaboration],
  Phys.\ Lett.\ B {\bf 739}, 229 (2014)
  [arXiv:1408.0806 [hep-ex]].
  
\bibitem{CMS:2015kda} 
  CMS Collaboration,
  CMS-PAS-EXO-12-043.

\bibitem{Bai:2014xba} 
  Y.~Bai and J.~Berger,
  arXiv:1407.4466 [hep-ph].
  
\bibitem{Chun:2014jha} 
  E.~J.~Chun, S.~Jung, H.~M.~Lee and S.~C.~Park,
  Phys.\ Rev.\ D {\bf 90}, no. 11, 115023 (2014)
  [arXiv:1408.4508 [hep-ph]].
  
  
\bibitem{Queiroz:2014pra} 
  F.~S.~Queiroz, K.~Sinha and A.~Strumia,
  Phys.\ Rev.\ D {\bf 91}, no. 3, 035006 (2015)
  [arXiv:1409.6301 [hep-ph]].
  
\bibitem{Allanach:2014nna} 
  B.~C.~Allanach, S.~Biswas, S.~Mondal and M.~Mitra,
  Phys.\ Rev.\ D {\bf 91}, no. 1, 015011 (2015)
  [arXiv:1410.5947 [hep-ph]].
  
\bibitem{Allanach:2015ria} 
  B.~Allanach, A.~Alves, F.~S.~Queiroz, K.~Sinha and A.~Strumia,
  arXiv:1501.03494 [hep-ph].
  

  
\bibitem{Varzielas:2015iva} 
  I.~de Medeiros Varzielas and G.~Hiller,
  JHEP {\bf 1506}, 072 (2015)
  [arXiv:1503.01084 [hep-ph]].
  
  
\bibitem{Belyaev:2005ew} 
  A.~Belyaev, C.~Leroy, R.~Mehdiyev and A.~Pukhov,
  JHEP {\bf 0509}, 005 (2005)
  [hep-ph/0502067].
  
\bibitem{Collaboration:2011qaa} 
  F.~D.~Aaron, C.~Alexa, V.~Andreev, S.~Backovic, A.~Baghdasaryan, S.~Baghdasaryan, E.~Barrelet and W.~Bartel {\it et al.},
  Phys.\ Lett.\ B {\bf 704}, 388 (2011)
  [arXiv:1107.3716 [hep-ex]];
  
\bibitem{Abramowicz:2012tg} 
  H.~Abramowicz {\it et al.}  [ZEUS Collaboration],
  Phys.\ Rev.\ D {\bf 86}, 012005 (2012)
  [arXiv:1205.5179 [hep-ex]].

\bibitem{Kramer:2004df} 
  M.~Kramer, T.~Plehn, M.~Spira and P.~M.~Zerwas,
  Phys.\ Rev.\ D {\bf 71}, 057503 (2005)
  [hep-ph/0411038].
  
\bibitem{Mandal:2015lca} 
  T.~Mandal, S.~Mitra and S.~Seth,
  arXiv:1506.07369 [hep-ph].
  
\bibitem{Blumlein:1992ej} 
  J.~Blumlein and R.~Ruckl,
  Phys.\ Lett.\ B {\bf 304}, 337 (1993);
  
  \bibitem{Hewett:1997ce} 
  J.~L.~Hewett and T.~G.~Rizzo,
  Phys.\ Rev.\ D {\bf 56}, 5709 (1997)
  [hep-ph/9703337].
  
\bibitem{Alwall:2014hca} 
  J.~Alwall, R.~Frederix, S.~Frixione, V.~Hirschi, F.~Maltoni, O.~Mattelaer, H.-S.~Shao and T.~Stelzer {\it et al.},
  JHEP {\bf 1407}, 079 (2014)
  [arXiv:1405.0301 [hep-ph]].
  
\bibitem{Sjostrand:2006za} 
  T.~Sjostrand, S.~Mrenna and P.~Z.~Skands,
  JHEP {\bf 0605}, 026 (2006)
  [hep-ph/0603175].
  
\bibitem{deFavereau:2013fsa} 
  J.~de Favereau {\it et al.}  [DELPHES 3 Collaboration],
  JHEP {\bf 1402}, 057 (2014)
  [arXiv:1307.6346 [hep-ex]].
  
\bibitem{Pumplin:2002vw} 
  J.~Pumplin, D.~R.~Stump, J.~Huston, H.~L.~Lai, P.~M.~Nadolsky and W.~K.~Tung,
  JHEP {\bf 0207}, 012 (2002)
  [hep-ph/0201195].
   
   
 \bibitem{MMSLQSNLO}
  T.~Mandal, S.~Mitra and S.~Seth, work in progress.   
   
\bibitem{Alves:2002tj} 
  A.~Alves, O.~Eboli and T.~Plehn,
  Phys.\ Lett.\ B {\bf 558}, 165 (2003)
  [hep-ph/0211441].
  
\bibitem{Hammett:2015sea} 
  J.~B.~Hammett and D.~A.~Ross,
  arXiv:1501.06719 [hep-ph].
  
     

  
\bibitem{Alwall:2008qv} 
  J.~Alwall, S.~de Visscher and F.~Maltoni,
  JHEP {\bf 0902}, 017 (2009)
  [arXiv:0810.5350 [hep-ph]].
  

  

\bibitem{Mandal:2012rx} 
  T.~Mandal and S.~Mitra,
  Phys.\ Rev.\ D {\bf 87}, no. 9, 095008 (2013)
  [arXiv:1211.6394 [hep-ph]].
  

\bibitem{Dorsner:2014axa} 
  I.~Dorsner, S.~Fajfer and A.~Greljo,
  JHEP {\bf 1410}, 154 (2014)
  [arXiv:1406.4831 [hep-ph]].
    

\end{thebibliography}
\end{document}